\documentclass{aa}  
\usepackage{natbib,twoopt}
\usepackage[breaklinks=true]{hyperref} 
\bibpunct{(}{)}{;}{a}{}{,} 
\makeatletter
   \newcommandtwoopt{\citeads}[3][][]{\href{http://adsabs.harvard.edu/abs/#3}%
    {\def\hyper@linkstart##1##2{}%
     \let\hyper@linkend\@empty\citealp[#1][#2]{#3}}}
   \newcommandtwoopt{\citepads}[3][][]{\href{http://adsabs.harvard.edu/abs/#3}%
    {\def\hyper@linkstart##1##2{}%
     \let\hyper@linkend\@empty\citep[#1][#2]{#3}}}
   \newcommandtwoopt{\citetads}[3][][]{\href{http://adsabs.harvard.edu/abs/#3}%
    {\def\hyper@linkstart##1##2{}%
     \let\hyper@linkend\@empty\citet[#1][#2]{#3}}}
   \newcommandtwoopt{\citeyearads}[3][][]%
    {\href{http://adsabs.harvard.edu/abs/#3}
    {\def\hyper@linkstart##1##2{}%
     \let\hyper@linkend\@empty\citeyear[#1][#2]{#3}}}
\makeatother
\usepackage{lscape}
\usepackage{romannum}
\usepackage{placeins}
\usepackage{subfig}
\usepackage{xcolor}
\usepackage{comment}
\AtBeginDocument{\pagenumbering{arabic}}

\usepackage{graphicx}
\usepackage{txfonts}
 \newcommand{\logg}{$\log g$}

 \newcommand{\starA}{\rm Gl~725A\,}

\begin{document}

   \title{Gl\,725A b: a potential super-Earth detected with SOPHIE and SPIRou in an M dwarf binary system at 3.5 pc}
\titlerunning{Gl\,725A b: a potential super-Earth}

  \author{P. Cort\'es-Zuleta \inst{\ref{inst},\ref{inst1},\thanks{Corresponding author: pcz1@st-andrews.ac.uk}} 
          \and
          I. Boisse \inst{\ref{inst1}}
          \and
         M. Ould-Elhkim \inst{\ref{inst2}}
         \and
         T. G. Wilson \inst{\ref{inst3}} 
         \and
         P. Larue \inst{\ref{inst7}}
         \and
         A. Carmona \inst{\ref{inst7}}
         \and
         X. Delfosse \inst{\ref{inst7}}
         \and
         J.-F. Donati \inst{\ref{inst2}}  
         \and
         T. Forveille \inst{\ref{inst7}}
         \and
         C. Moutou \inst{\ref{inst2}}
         \and
         A. Collier Cameron \inst{\ref{inst}}
         \and
         \'E. Artigau \inst{\ref{inst8}}
         \and
         L. Acuña\inst{\ref{inst1},\ref{MPIA}}
         \and
         L. Altinier\inst{\ref{inst1}}
         \and
         N. Astudillo-Defru\inst{\ref{inst21}}
         \and
         C. Baruteau\inst{\ref{inst2}}
         \and
         X. Bonfils\inst{\ref{inst7}}
         \and
         S. Cabrit\inst{\ref{inst20}}
         \and
         C. Cadieux\inst{\ref{inst8}}
         \and
         N. J. Cook\inst{\ref{inst8}}
         \and
         E. Decocq\inst{\ref{inst1}}
         \and
         R. F. D\'iaz\inst{\ref{inst9}}
         \and
         P. Fouqu\'e\inst{\ref{inst2}}
         \and
         J. Gomes da Silva\inst{\ref{inst10}}
         \and
         K. Grankin\inst{\ref{inst16}}
         \and
         S. Grouffal\inst{\ref{inst1}}
         \and
         N. Hara\inst{\ref{inst1}}
        \and
         G. H\'ebrard\inst{\ref{inst11},\ref{inst13}}
         \and
         N. Heidari\inst{\ref{inst11}}
         \and
         J. H. C. Martins\inst{\ref{inst10}}
         \and
         E. Martioli\inst{\ref{inst4},\ref{inst11}}
         \and
         M. Maurice\inst{\ref{inst23}}
        \and
        J. Scigliuto\inst{\ref{inst15}}
        \and
        J. Serrano Bell\inst{\ref{inst9}}
        \and
        S. Sulis\inst{\ref{inst1}}
        \and
        A. C. Petit\inst{\ref{inst15}}
        \and
        H. G. Vivien\inst{\ref{inst1}}
                 }

   \institute{\label{inst}Centre for Exoplanet Science, SUPA, School of Physics and Astronomy, University of St Andrews, North Haugh, St Andrews KY16 9SS, UK
   \and
   \label{inst1} Aix Marseille Univ, CNRS, CNES, LAM, Marseille, France 
                 \and
            \label{inst2} Univ. de Toulouse, CNRS, IRAP, 14 av. Belin, 31400 Toulouse, France
            \and
            \label{inst3} Department of Physics, University of Warwick, Gibbet Hill Road,
            Coventry CV4 7AL, United Kingdom
            \and
        \label{inst7} Univ. Grenoble Alpes, CNRS, IPAG, 38000 Grenoble, France
         \and
            \label{inst8} Institut Trottier de Recherche sur les Exoplanètes and Département de Physique, Université de Montréal, 1375 Avenue Thérèse-Lavoie-Roux, Montréal, QC, H2V 0B3, Canada
            \and
            \label{MPIA} Max-Planck Institut für Astronomie, Königstuhl 17, 69117 Heidelberg, Germany
             \and
          \label{inst21} Departamento de Matem\'atica y F\'isica Aplicadas, Universidad Cat\'olica de la Sant\'isima Concepci\'on, Alonso de Rivera 2850, Concepci\'on, Chile
          \and
        \label{inst20} Observatoire de Paris, CNRS, LERMA, Sorbonne Université, 61 avenue de l’Observatoire, 75014 Paris, France
          \and
          \label{inst9} International Center for Advanced Studies (ICAS) and ICIFI (CONICET), ECyT-UNSAM, Campus Miguelete, 25 de Mayo y Francia, (1650) Buenos Aires, Argentina.
          \and
          \label{inst10} Instituto de Astrof \'isica e Ci\^encias do Espaço, Universidade do Porto, CAUP, Rua das Estrelas, 4150-762 Porto, Portugal
          \and
        \label{inst16} Crimean Astrophysical Observatory, 298409 Nauchny, Republic of Crimea
          \and
          \label{inst11} Institut d'Astrophysique de Paris, CNRS, UMR 7095, Sorbonne Universit\'{e}, 98 bis bd Arago, 75014 Paris, France
          \and
          \label{inst13} Observatoire de Haute-Provence, CNRS, Universit\'e d'Aix-Marseille, 04870 Saint-Michel-l'Observatoire, France
          \and
          \label{inst4} Laborat\'orio Nacional de Astrof\'isica, Rua Estados Unidos 154, 37504-364, Itajub\'a - MG, Brazil 
          \and
          \label{inst23} Laboratoire de Météorologie Dynamique, Institut Pierre Simon Laplace, CNRS, Paris, France
          \and
        \label{inst15} Université Côte d'Azur, Observatoire de la Côte d'Azur, CNRS, Laboratoire Lagrange, France
}

   \date{Received ; accepted }

 
  \abstract
{We report the discovery of a super-Earth candidate orbiting the nearby mid M dwarf Gl\,725A using the radial velocity (RV) method. The planetary signal has been independently identified using high-precision RVs from the SOPHIE and SPIRou spectrographs, in the optical and near-infrared domains, respectively. We modelled the stellar activity signal jointly with the planet using two Gaussian Processes, one for each instrument to account for the chromaticity of the stellar activity and instrumental systematics, along with a Keplerian model. The signal is significantly detected with a RV semi-amplitude of $1.67\pm0.20$ m/s. The planet Gl\,725A b is found to be in an orbit compatible with circular with a period of $11.2201\pm0.0051$ days. We analysed 27 sectors of TESS photometry on which no transit event was found. We determined a minimum mass of $M_{p}\sin{i}=2.78\pm0.35\,M_{\oplus}$ which places the planet in the super-Earth regime. Using Mass-Radius relationships we predict a planetary radius to be between 1.2 and $2.0\,R_{\oplus}$. The proximity of Gl\,725A, of only 3.5\,pc, makes this new exoplanet one of the closest to Earth and joins the group of S-type low-mass planets in short orbits ($P<15$ d) around close M dwarfs. 
}


   \maketitle
%

\section{Introduction}


M dwarfs are cool and low-mass stars that comprise the most abundant stellar type in our Galaxy, reaching up to 75\% of the galactic stellar population \citep[]{Henry2006}. They are attractive targets for exoplanetary science, in particular for the search of Earth-mass planets. Due to their low mass and size, the Doppler shift and transit depth produced by a small planet orbiting an M dwarf is larger than that of a Sun-like star, for a given planet mass/radius, and orbital distance.

\citet{Reyle2021} found that 61$\%$ of stars and brown dwarfs within the Solar neighborhood (d<10\,pc) are M dwarfs. However, only 24$\%$ of these group of close M dwarfs have known exoplanets\footnote{As of March 2024, source: the Exoplanet Archive \url{exoplanetarchive.ipac.caltech.edu}}. Most of the exoplanets around M dwarfs in the Solar neighbourhood have been discovered with the radial velocity (RV) technique and therefore, only a minimum mass can be determined. Regarding the multiplicity of the stellar host, \citet{Reyle2021} found a stellar multiplicity frequency of 27$\%$ in the Solar neighbourhood. However, only nine multiple M dwarf systems have been confirmed as exoplanet hosts, and for all the them, the planets found only orbit one component of the system. This type of planet is known as circumstellar or S-type planet \citep{Dvorak1986}. 

Several surveys focused on searching for planets around M dwarfs have started in the last decades, using transit photometry and RV techniques. On the photometry side, some examples of projects are MEarth \citep{Nutzman2008}, TRAPPIST \citep{Gillon2011}, SPECULOOS \citep{Delrez2018}, and TIERRAS \citep{Garcia2020}. On the other hand, projects such as: HPF \citep{Mahadevan2012}, HARPS \citep{Bonfils2013}, HARPS-N \citep{Covino2013}, IRD \citep{Kotani2014}, CARMENES \citep{Quirrenbach2018}, MAROON-X \citep{Seifahrt2018}, SOPHIE \citep{Hobson2018}, SPIRou \citep{Donati2020}, and NIRPS \citep{Artigau2024}, rely on high-precision Doppler spectroscopy. One key results of these surveys is that low-mass planets (1 < $M_{p}\sin{i}$/$M_{\oplus}$ < 10) are common around M dwarfs \citep{Bonfils2013, Kopparapu2013, Dressing2013, Dressing2015, Sabotta2021} and their frequency may be closely related to the spectral type \citep[e.g.,][]{Pinamonti2022}.

However, despite the technical developments and efforts to discover low-mass planets around M dwarfs, stellar activity is currently one of the main challenges that hinders their detection. Cool stars host magnetic fields that generate activity features all over the stellar surfaces \citep{Reiners2012}. These features co-rotate with the star producing quasi-periodic variations in the observed RVs. These RV variations induced by stellar activity can reach amplitudes of a few meters per second, even for less active stars, which can lead to false-positive exoplanet detection \citep[e.g.,][]{Queloz2001,Huelamo2008,Carolo2014,Bortle2021}.

The development of near-infrared (NIR) spectrographs has been fundamental to enriching our understanding of the activity of M dwarfs. These stars emit most of their flux at NIR wavelengths, allowing to achieve higher sensitivity and precision than in the optical domain. Moreover, the spot-induced RV variability tends to decrease as a function of wavelength \citep{Martin2006, Desort2007, Reiners2010, Mahmud2011,Cale2021} since the flux contrast between the stellar surface and the dark spots is smaller at NIR wavelengths, and therefore the induced RV jitter decreases \citep{Reiners2010,Andersen2015}.

As the stellar activity signal is chromatic, optical and NIR observations are of great help to disentangle the stellar activity jitter from the Keplerian signals \citep{Reiners2013, Robertson2020}. TW Hydrae \citep{Huelamo2008}, and AD Leo \citep{Carleo2020,Carmona2023} are two examples of how multi-wavelength observations helped to reject the planet candidates as activity-induced false positives. 

SOPHIE and SPIRou are high-precision spectrographs in the optical and NIR domains, respectively. Our team is carrying out a synergy program in which dozens of early to mid-M dwarfs are observed with both instruments, providing in many cases contemporary observations. This allows us to have a better phase coverage of planet candidates and to study the chromaticity of the stellar activity signatures in the RVs. The results from this program (e.g., \citealp{Carmona2023}, \citealt{CortesZuleta2023}) highlight the importance of using multi-wavelength observations to understand the activity of M dwarfs and their planets.



In this work, we present the analysis of Gl\,725 A, the brighter component of an M dwarf binary system in the Solar neighbourhood, located at only 3.5 pc from Earth. The mean angular separation of the binary system is of 14.9$^{\prime\prime}$ which allows gathering individual high-precision spectroscopy. Gl\,725 A is a mid M dwarf with low activity levels and a rotation period previously measured at $103.1\pm6.1$ d\,\citep{Donati2023}. The analysis of SOPHIE and SPIRou high-precision RV time series indicates the presence of a low-mass planet in a 11-day orbit, potentially in the super-Earth regime. This paper is structured as follows. Section \ref{sec:obs} details the observations and data reduction. Sections \ref{sec:stellar} and \ref{sec:activity} describe the stellar properties and activity analysis, respectively. The detection and characterization of the planet Gl\,725A b is detailed in Sect. \ref{sec:planet}. The limits of planet radius and the search for transits in TESS data is discussed in Sect. \ref{sec:tess}. Finally, discussions and conclusions are detailed in Sects. \ref{sec:discussion} and \ref{sec:conclusions}, respectively.


\begin{table}
\centering
\footnotesize
 \caption[]{\label{tab:stellarparams} Identifiers and stellar parameters of Gl\,725 A and Gl\,725 B.}
 
 
\begin{tabular}{l l | c c}
 \hline \hline
\multicolumn{4}{c}{Star A} \\
 \hline \hline
 Identifiers & & \\
 \hline
HD & & 173739 \\
Gliese & & 725 A \\
2MASS & & J18424666+5937499  \\
Karmn & & J18427+596N\\
TIC & & 359676790 \\
Gaia DR3 & & 2154880616774131840\\

 \hline
  Parameter &
  Units&
  Value &
  References \\
 \hline

 RA  & h m s &  18 42 46.704386 & 1\\
 dec & d m s &  +59 37 49.409446 & 1 \\
 pmRA & mas/yr & $-1311.679\pm0.027$  & 1 \\
 pmdec & mas/yr & $1792.325\pm0.026$ & 1\\
 Parallax & mas &  $283.84 \pm 0.02$ & 1\\ 
Distance & pc  & $3.5231 \pm 0.0002$ & 1\\
Spectral type & &  M3 & 2 \\
$V$ & mag & $8.93\pm0.02$ & 3 \\
$B$ & mag & $10.43\pm0.05$ & 3\\
$G$ & mag & $7.854\pm0.002$ & 1 \\
$J$ & mag & $5.19\pm0.02$ & 4 \\
$H$ & mag & $4.47\pm0.04$ & 4 \\
$K$ & mag & $4.43\pm0.02$ & 4 \\

Mass &\(M_\odot\) & $ 0.330\pm0.008 $ & this work \\ 
Radius &\(R_\odot\) & $0.351 \pm0.003 $ & this work \\
$\log \rm (L_{*} / L_\odot)$& & $-1.809\pm0.002$ & 5\tablefootmark{a}\\

$P_{\rm rot}$& d & $103.1\pm6.1$  & 6 \\ 
\logg$_{\rm ~SOPHIE}$ & dex & $4.90 \pm 0.33$ & this work\tablefootmark{b}\\
\logg$_{\rm ~SPIRou}$ & dex & $4.77 \pm 0.06$ &5\tablefootmark{a}\\
\logg & dex & $4.87\pm0.01$ & this work\tablefootmark{c} \\
$ \rm T_{eff,~SOPHIE}$& K & $3433\pm68$& this work\tablefootmark{b} \\
$ \rm T_{eff,~SPIRou}$& K & $3470\pm31$ & 5\tablefootmark{a}\\
$\rm [Fe/H]$ & dex  & $-0.19\pm0.13$ & this work\tablefootmark{b}\\
$\rm [M/H]$ & dex & $-0.26\pm0.10$ & 5\tablefootmark{a} \\
$\rm [\alpha/Fe]$ & dex & $0.15\pm0.04$ & 5\tablefootmark{a} \\
Age & Gyr & $6.2\pm2.3$\ & this work\tablefootmark{a} \\
\hline \hline
\multicolumn{4}{c}{Star B} \\
\hline \hline
 Identifiers & & \\
 \hline
HD & & 173740 \\
Gliese & & 725 B \\
2MASS & & J18424688+5937374  \\
Karmn & & J18427+596S\\
TIC & & 392572237 \\
Gaia DR3 & & 2154880616774131712\\
\hline
Spectral type & & M4 & 2 \\
Mass &\(M_\odot\)  & $0.25\pm0.02$ & 5\\
Radius &\(R_\odot\)  & $0.280\pm0.005$ & 5\\
$\log \rm (L_{*} / L_\odot)$ & & $-2.038\pm0.003$ & 5\\
$P_{\rm rot}$& d & $135\pm15$ & 6\\
$ \rm T_{\rm eff}$& K & $3379\pm31$& 5 \\
\hline \hline

\end{tabular}
\tablebib{(1)~\citealt{GaiaDR3}; (2)~\citealt{Fouque2018}; (3)~\citealt{Fabricius2002}; (4)~\citealt{Cutri2003}; (5)~\citealt{Cristofari2022b};  (6)~\citealt{Donati2023}}
\tablefoot{
\tablefoottext{a}{Derived from SPIRou spectra.}
\tablefoottext{b}{Derived from SOPHIE spectra.}
\tablefoottext{c}{Derived from the estimated mass and radius.}
}
\end{table}


\section{Observations and data reduction}\label{sec:obs}

\subsection{SOPHIE radial velocities}

SOPHIE (Spectrographe pour l’Observation des Phénomènes des Intérieurs stellaires et des Exoplanètes) is a high-resolution, fiber-fed, échelle spectrograph mounted on the 1.93m telescope at the Observatoire de Haute-Provence (OHP, \citealt{Perruchot2008}). The SOPHIE wavelength domain covers an optical range from 390 nm to 690 nm across 39 spectral orders. The instrument has two observation modes: high-efficiency (HE, R=40\,000) and high-resolution (HR, R=70\,000). The instrumental drift is measured by performing simultaneous calibrations with a super-continuum lamp filtered through a Fabry-Pérot (FP) etalon.

The star Gl\,725A was observed in the HR mode between March 2021 and December 2023 as part of the subprogram 3 (SP3) of the SOPHIE exoplanet consortium. This program is dedicated to the search of exoplanets around M dwarfs (e.g., \citealp{Hobson2018}, \citealt{Hobson2019}, \citealp{Diaz2019}), and the study of the chromaticity of their stellar activity signal (e.g., \citealp{CortesZuleta2023,Carmona2023}). The SOPHIE aperture diameter is 3$^{\prime\prime}$ and therefore, we can discard contamination from the stellar companion. A total of 176 observations were collected, of which 12 were removed due to low signal-to-noise (S/N<30), high airmass (>1.6), bad weather conditions, or technical problems. The mean exposure time of the remaining observations is 970 s and their mean SNR is 72 per pixel at 550 nm. 

The extraction and reduction of the spectra were done using the SOPHIE Data Reduction Software (DRS, \citealp{Bouchy2009a,Heidari2024}). The DRS performs spectral order localization, optimal order extraction, cosmic-ray rejection, and wavelength calibration. The spectra are also corrected for the charge transfer inefficiency (CTI) effect on the CCD following \citet{Bouchy2009a} and \citet{Hobson2018}. The DRS applies the cross-correlation function (CCF) technique to derive the RVs by cross-correlating an empirical M5V mask to the observed spectra. 

However, the CCF method is not optimal for deriving high-precision RVs from M dwarf spectra because of the large number of blended absorption lines. Therefore, we applied the NAIRA code \citep{Astudillo2017b} which is based on the template matching technique, to use the most of the spectral information available. In this method, all the available spectra are used to build a high-S/N stellar template. The measured RV is the minimum of the $\chi^2$ profile of the residuals between the individual observed spectrum and the shifted stellar template. From NAIRA we also obtained standard activity indicators based on the CCF such as the BIS, FWHM, and contrast, and spectral activity indices based on the H$\alpha$ line and the Ca \Romannum{2} H\&K lines (S index).

The RV zero-point variations of the instrument \citep{Courcol2015} are corrected using a RV zero-point correction time series built from nightly observations of a list of standard stars, consisting of a group of G-type stars: HD\,185144, HD\,9407, and HD\,89269A.

\begin{figure}
  \centering
   \includegraphics[width=0.95\linewidth]{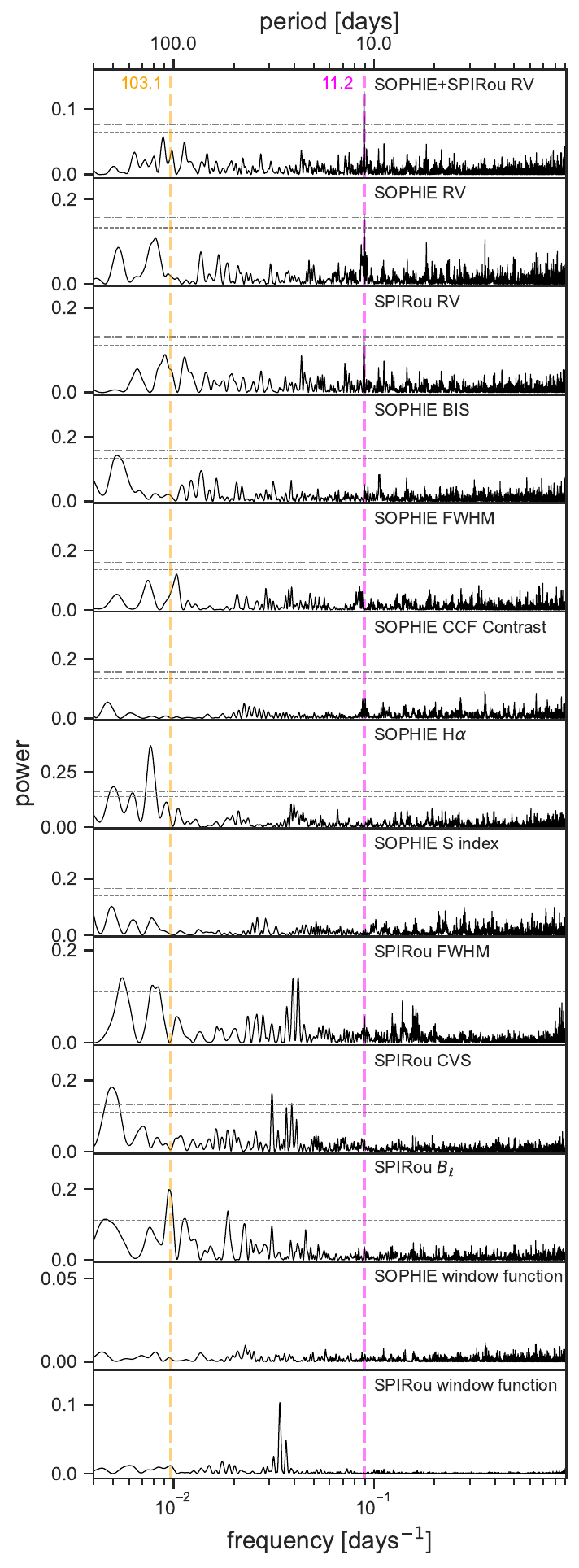}
  \caption{Periodograms of the SOPHIE and SPIRou RVs and activity indicators. The pointed and dashed horizontal lines mark the 0.1 and 1$\%$ FAP. The magenta and orange dashed vertical lines depict the orbital period of the planet at 11.2 d and the stellar rotation period at 103.1 d, respectively.}\label{fig:periodograms}%
   \end{figure}

The final RVs computed by NAIRA have an average uncertainty of 1.7\,m/s with a scatter of 3.1\,m/s, and they are listed in Table \ref{table:SOPHIEdata}, along with the activity indicators described in Sect. \ref{sec:activity}. We computed the Generalized Lomb-Scargle periodogram to find the periodicities in the RV time series\footnote{Throughout this work we computed the Generalized Lomb-Scargle periodogram using the \texttt{astropy} package \citep{astropy} and their corresponding false alarm probability (FAP) is computed following \citealt{Baluev2008}.}. We found a significant periodicity at 11.2 days with a $\log(\rm FAP)=-2.2$ (see Fig. \ref{fig:periodograms}).

\subsection{SPIRou radial velocities}

\subsubsection{APERO reduction}
SPIRou (Spectro-Polarimetre InfraRouge, \citealp{Donati2020}) is a high-resolution spectropolarimeter and velocimeter mounted at the Canada-France-Hawaii Telescope (CFHT). It covers the near-infrared spectral domain from 980 nm to 2400 nm, which corresponds to the \textit{Y}, \textit{J}, \textit{H}, and \textit{K} bands, across 49 spectral orders. The spectral resolution of SPIRou is $\lambda/\Delta\lambda \sim 70\,000$. Each SPIRou observation consists of a spectropolarimetric sequence of four sub-exposures, each one with a different rotation angle of the half-wave Fresnel rhombs in the polarimeter.

The observations of Gl\,725A were carried out between February 2019 and April 2022 as part of the Planet Search program (WP1, \citealt{Moutou2023}) of the SPIRou Legacy Survey (SLS, \citealt{Donati2020}, whose main goal is to perform a systematic RV monitoring of nearby M dwarfs. The SPIRou on sky diameter of circular instrument aperture is 1.29$^{\prime\prime}$, thus we can discard possible contamination from the stellar companion. A total of 213 four-sub-exposures sequences were collected, with a median exposure time of 61 s per rhomb position (244 s for a complete sequence), and a median S/N per pixel at 1670 nm of 201. We removed 12 sequences due to low S/N (S/N<150) or high airmass (>1.7).

We used the SPIRou data reduction software APERO v.0.7.275\citep{Cook2022} to extract the spectra. APERO performs an automatic reduction of the raw images, including correction for detector effects, and identification and removal of bad pixels and cosmic rays. The data undergo calibration through flat and blaze corrections, followed by optimal extraction \citep{Horne1986} from both science channels (fibers A and B carrying orthogonal polarimetric states of the incoming light) and the calibration channel (fiber C). The pixel-to-wavelength calibration is performed following \citet{Hobson2021}, using an UNe hollow cathode lamp and an FP etalon. 

APERO conducts a two-step process for telluric and night emission correction. In the initial step, an atmospheric absorption model is generated using TAPAS (Transmissions of the AtmosPhere for AStronomical data; \citealt{Bertaux2014}). This model is substracted from the science frames, leaving only percent-level residuals in deep (>50\%) lines of H$_2$O and dry absorption molecules (CH$_4$, O$_2$, CO$_2$, N$_2$O, and O$_3$). The process of constructing a TAPAS model is also implemented for a series of rapidly rotating hot stars observed under various atmospheric conditions to establish a library of telluric residual models. This library of telluric residual models has three degrees of freedom (optical depths of H$_2$O and dry components, and a constant). In the second step, the telluric residual model is subtracted from the data to derive telluric-corrected spectra.

We obtained high-precision RVs using the line-by-line (LBL) method based on the framework proposed by \citet{Dumusque2018} and optimized for SPIRou data by \citet{Artigau2022}. The LBL method exploits the RV information per line in the spectra to generate a single RV measurement, calculated based on a simple mixture model that calculate a probability that a line is valid or not, giving less weight to the invalid lines in the weighted average calculation. The RV uncertainty is determined according to the methodology outlined by \citet{Bouchy2001}. This approach is extended to the calibration fiber to correct the instrumental drift. The RVs from the LBL are further corrected for instrumental drift and long-term zero point using a Gaussian Process Regression (GPR) with data from the most frequently observed stars in the SPIRou Legacy Survey, in a similar procedure as for SOPHIE \citep{Cadieux2022}.


\subsubsection{\texttt{Wapiti} RV correction}\label{wapiti}
One particular challenge of NIR spectroscopy is the contamination by telluric lines in the stellar spectra due to the Earth's motion around the Sun. These absorption and emission lines contaminate the stellar spectra and consequently, the computation of high-precision RVs is compromised \citep{Bean2010,Merwan2023}.

The SPIRou NIR spectra of Gl\,725A are importantly contaminated by telluric lines due to its high ecliptic latitude and therefore narrow Barycentric Earth RV (BERV) coverage, which ranges between $-5$ and $+5$\,km/s. To correct for this effect, we applied the \texttt{Wapiti} method (Weighted-principAl comPonent analysIs reconsTructIon; \citealt{Merwan2023}). This method is a data-driven approach
developed not only to correct for the effect of telluric lines but that from any systematics affecting the RVs, such as persistence in infrared arrays, glitches due to detector anomalies, and unidentified sources. This correction is only applied in the SPIRou spectra since telluric contamination is not that important in the optical domain, and furthermore, telluric regions are excluded in the SOPHIE RV computation.

\begin{figure}
\centering
\includegraphics[width=0.9\linewidth]{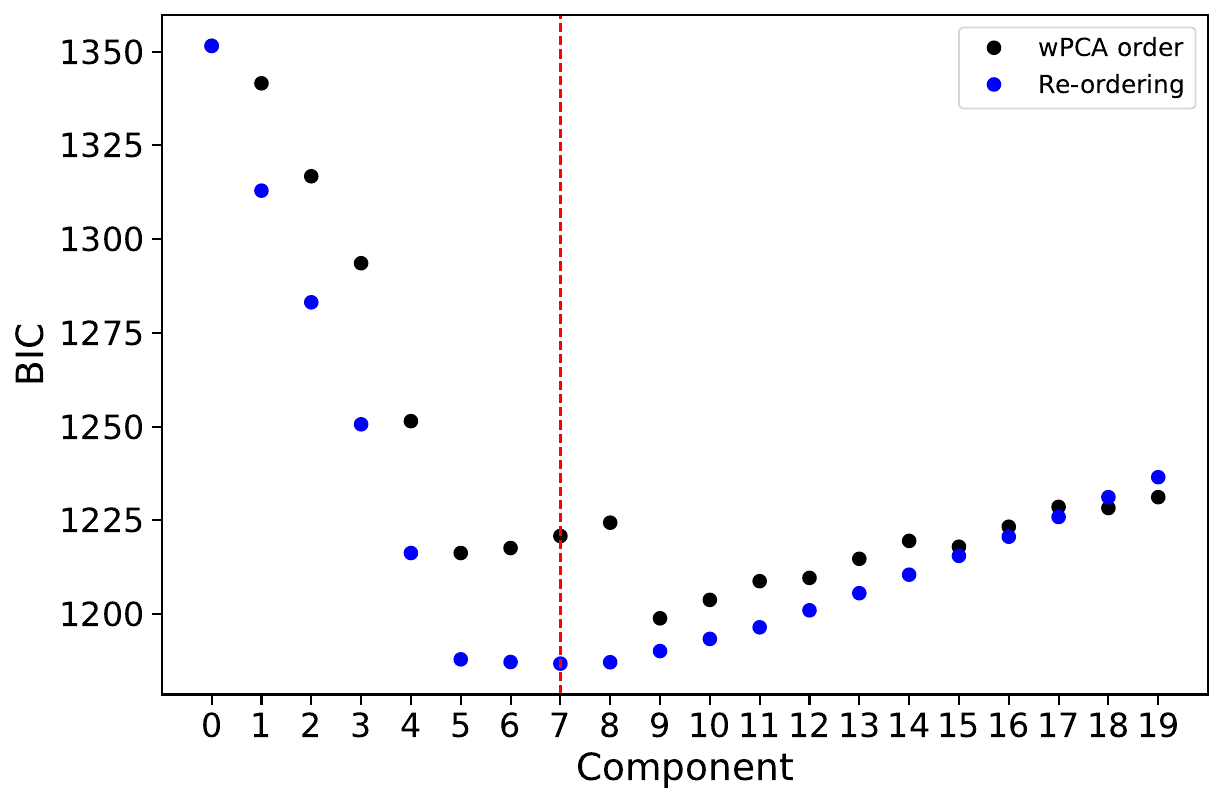}
\caption{Re-order of the wPCA components in decreasing BIC. The black circles correspond to the original order and in blue are re-ordered. The minimum BIC is found with 7 components, marked with the vertical dashed red line.}\label{fig:bic_reorder}
\end{figure}

In summary, \texttt{Wapiti} applies the weighted principal component analysis (wPCA, \citealt{Delchambre2015}) in the LBL per-line RV time series. A first selection of components is found through a permutation test by independently shuffling the columns of the data set and then computing the wPCA in the remaining data. This procedure is performed 100 times, where the variance of the permuted data is compared with the original variance. Relevant components are expected not to increase the variance of the data set. In the case of Gl\,725A we found that only the 19 first components cannot be explained by noise. The significant of these components is assessed by leave-p-out cross-validation (see e.g., \citealt{Merwan2023} and \citealt{Cretignier2023}) to measure the stability of each component to removing certain lines. Components that show excessive sensitivity to line removal are discarded. We found that only 15 components are chromatically stable. The final number of components is found through re-ordering the components depending on their Bayesian Information Criteria (BIC). The idea of this process is to find the number of component that minimizes the BIC. After this procedure, we found that 7 components is the optimal number to re-construct the systematics of the Gl\,725A data, as shown in Fig. \ref{fig:bic_reorder}. These components have periodicities of $\sim365$, $\sim180$, and $\sim90$ days and are correlated with the BERV, which indicate their origin from systematics, possibly telluric lines. The corrected RVs are computed as the difference between the original time series and the reconstructed time series using the optimal number of principal components. More details of this methodology will be given in Ould-Elhkim et al. (in prep).

\begin{figure}
\centering
\includegraphics[width=\linewidth]{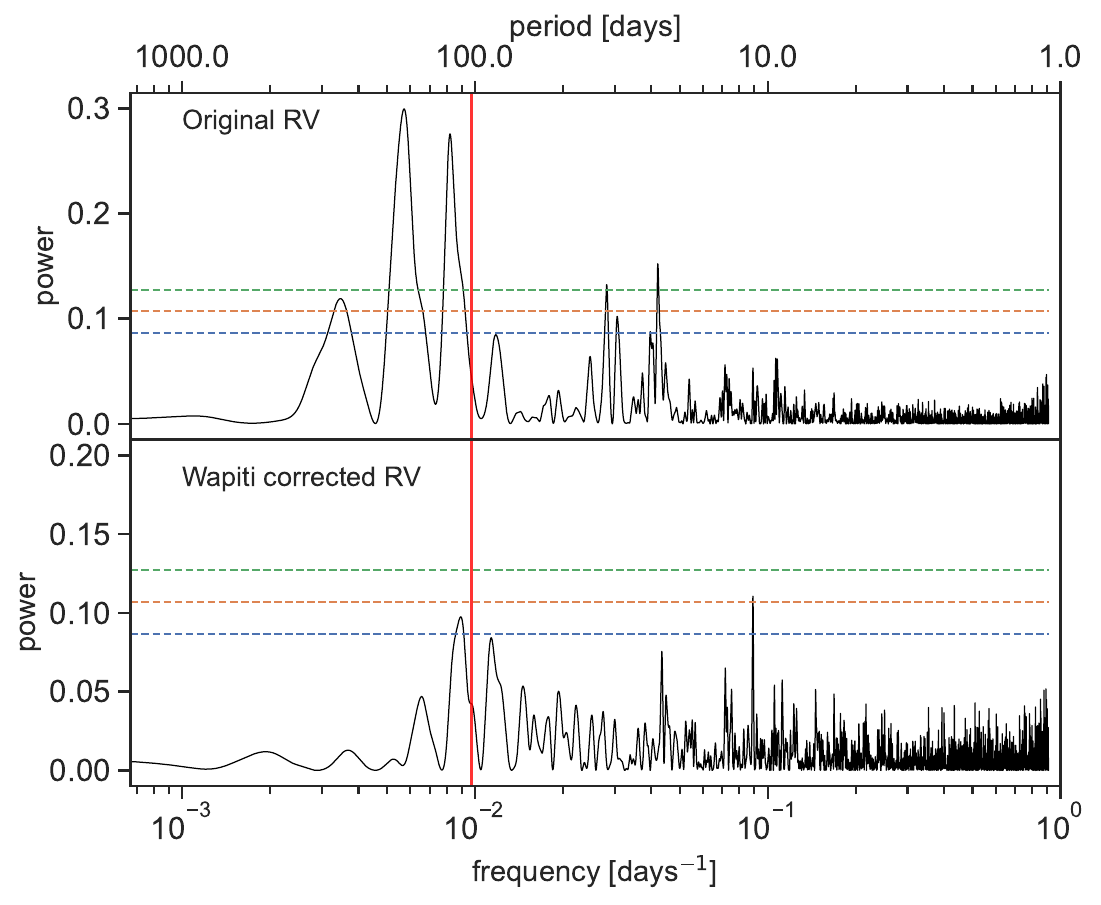}
\caption{Periodograms of the original SPIRou RVs (top) and after the \texttt{wapiti} correction (bottom), on which the period with the highest power correspond to the planet signal at 11.2 days. The horizontal dashed lines in blue, orange, and green depict the FAP at 10, 1, and 0.1$\%$, respectively. The vertical line indicates the stellar rotation period at 103.1 days. Details odf the \texttt{wapiti} correction can be found in Sect. \ref{wapiti}.}\label{fig:wapiti_periodograms}
\end{figure}

  As seen in the top panel of Fig.~\ref{fig:wapiti_periodograms}, the original SPIRou RVs show long-term periodicities, on which the highest power is found at 174\,days. This period is longer than the expected stellar rotation period of $P_{\rm rot}=103.1\pm6.1$ d and is closer to half of the Earth's orbit ($\sim180$ days). After the \texttt{Wapiti} correction, we computed the Generalized Lomb-Scargle periodogram of the SPIRou RVs. The periodogram in Fig. \ref{fig:wapiti_periodograms} shows a significant signal at the period of 11.2 d with a $\log$(FAP)=-1.4. There is a second signal at 111.9 d but less significant with a $\log$(FAP)=-0.8. 

The observations are listed in Table \ref{table:SPIRoudata} which includes the
RV and their uncertainties. Each RV data point is the binning of the 4 polarimetric observations. The SPIRou RV time series has an average uncertainty of 1.4\,m/s and scatter of 3.8\,m/s.

\subsection{TESS photometry}

The Transiting Exoplanet Survey Satellite (TESS, \citealt{Ricker_2015}) has observed most of the sky looking for exoplanets transiting in front of bright stars. Due to its favorable position in the sky, the binary system Gl\,725 has been observed in a total of 28 TESS Sectors: 14 to 26, 40, 41, 47 to 57, 59, and 60, covering a period between July 2018 and January 2023. Sector 14 was excluded from the analysis as the star was too close to a detector edge. The TESS pixel size is 21$^{\prime\prime}$ but 
 the mean angular separation between Gl 725 A and B is 14.9$^{\prime\prime}$, consequently both stars are either in the same or in contiguous pixels.

\begin{figure*}
\centering
\includegraphics[width=0.99\linewidth]{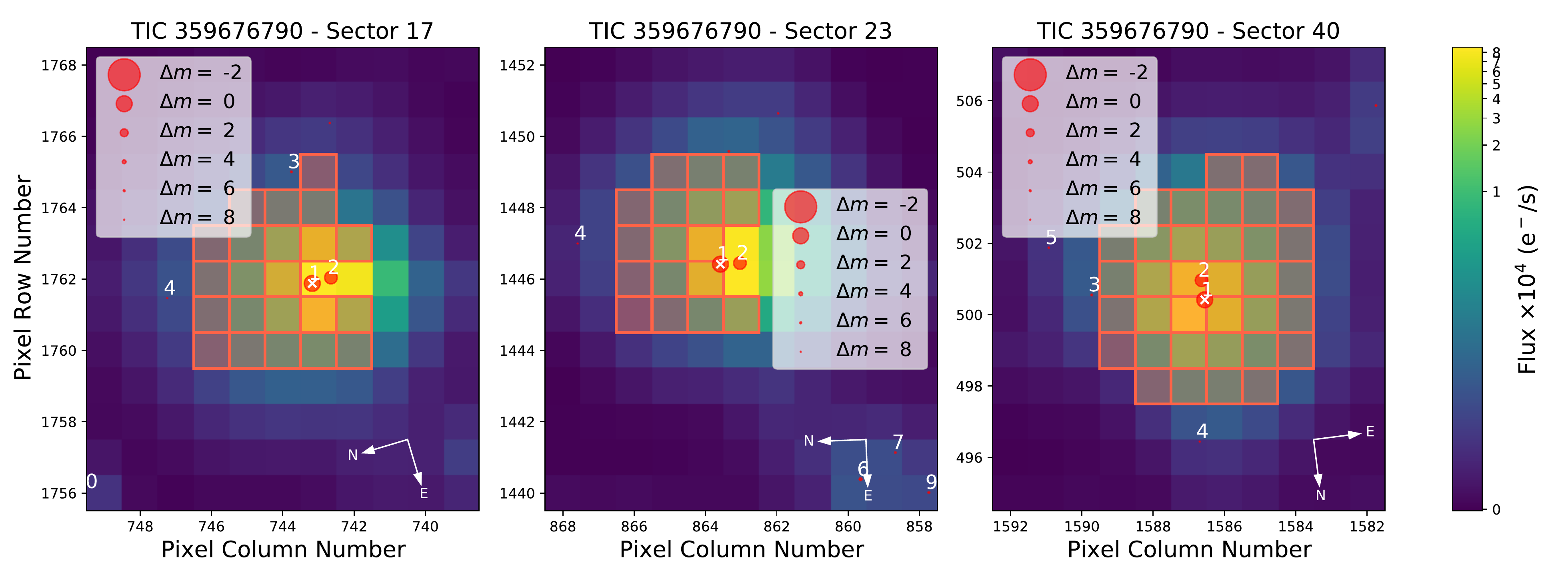}
\caption{TESS target pixel file images of Sectors 17, 23, and 40, generated with \texttt{tpfplotter}. These sectors are shown as example of the aperture photometry performed by the SPOC pipeline, where the inclusion of the stellar companion Gl 725 B is not consistent.}\label{fig:TESS_sectors}
\end{figure*}

We obtained the Presearch Data Conditioning (PDC) flux processed by the TESS Science Processing Operations Center (SPOC). However, the aperture used to perform the flux extraction is not consistent across the sectors, meaning that in some sectors the aperture does not include both components of the system. Some examples of the target pixel file images generated with the software \texttt{tpfplotter}\footnote{\url{https://github.com/jlillo/tpfplotter}}\citep{Aller2020} are shown in Fig. \ref{fig:TESS_sectors}. Due to the proximity between the binary components and their similar magnitudes, it is not possible to independently extract their fluxes. For the sake of consistency, we performed a custom aperture photometry extraction of the fluxes including Gl 725 A and B using the package \texttt{lightkurve} \citep{lightkurve}. For each sector, an specific aperture was determined by eye to include only the binary components avoiding nearby sources. Nevertheless, contamination from nearby sources is not relevant for Gl\,725 since they are not close enough to be included in the aperture and are several magnitudes fainter, as seen in Fig. \ref{fig:TESS_sectors}.\\

\begin{figure}
\centering
\includegraphics[width=0.99\linewidth]{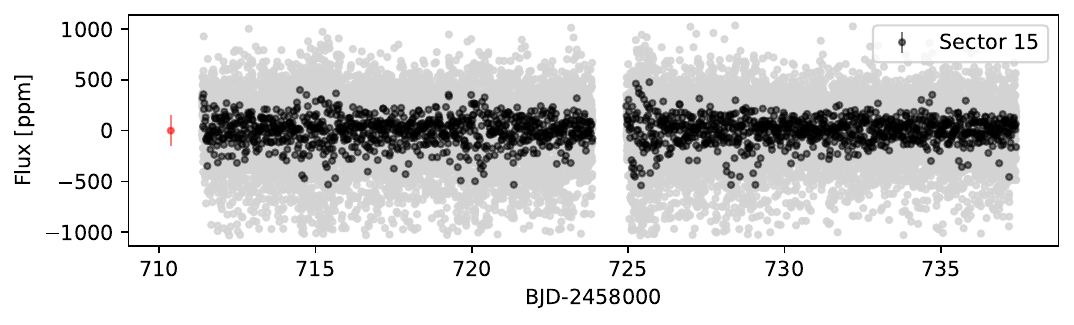}
\includegraphics[width=0.99\linewidth]{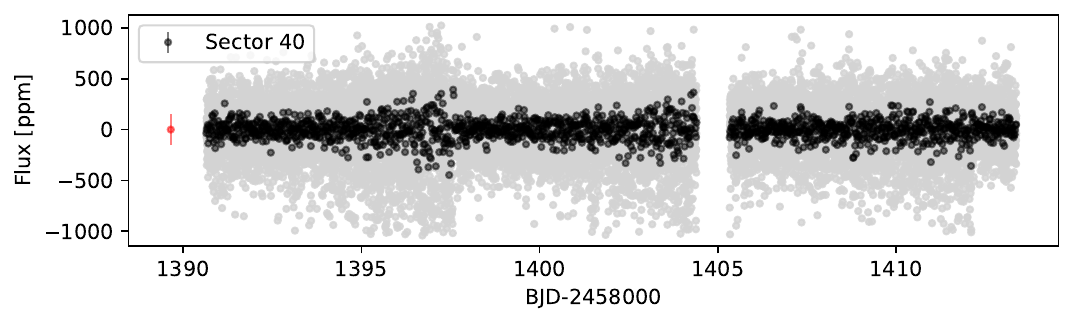}
\includegraphics[width=0.99\linewidth]{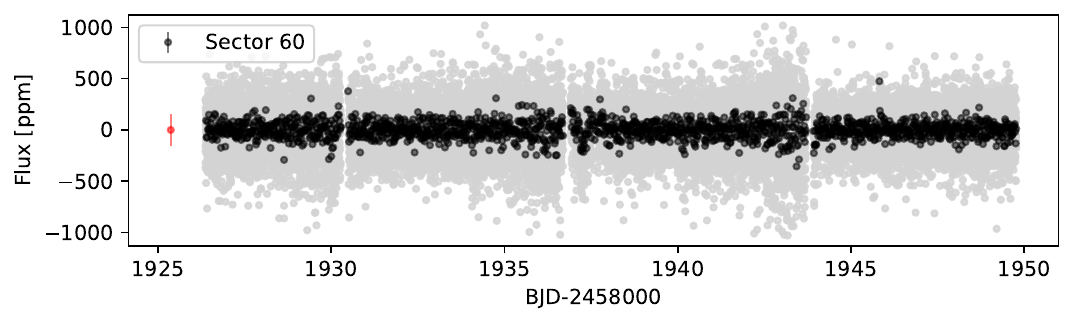}
\caption{Examples of the Gl\,725 detrended TESS light curves of sectors 15, 40, and 60. In light grey is the unbinned data and in black is the binned data in a 20 min window. To avoid overcrowding in the plots, the point in red shows the mean size of the error bars for the binned data.}\label{fig:TESS_lc_examples}
\end{figure}

The extracted light curves were cleaned of 3-$\sigma$ outliers and detrended using the \texttt{wotan}\footnote{\url{https://github.com/hippke/wotan}} Python package \citep{Hippke2019b}. This software includes several methods to perform light-curve detrending. We applied the method based on a time-windowed sliding filter with a robust location estimator, defined as "biweight'' in \texttt{wotan}. This is the method with best results in \citealt{Hippke2019b}. We applied a window size of 0.75 and we excluded the edges of the light curves, because these zones are usually affected by strong systematic errors. The detrended light curves have a scatter between 215 and 372 ppm, with a mean scatter of 290 ppm. Some examples of the light curves are shown in Fig. \ref{fig:TESS_lc_examples}.



\section{Stellar properties}\label{sec:stellar}

\starA is a bright (V $=8.93$ mag), M3V star at a distance of 3.5231\,pc \citep{GaiaDR3}. It is the brightest of a binary system on which the stellar companion, Gl\,725B, is a M3.5V star. Gl\,725A has a mass of $0.330\pm0.008\,$M$_{\odot}$ and a radius of $0.351\pm0.003$\,R$_{\odot}$. The main stellar parameters of \starA are listed in Table \ref{tab:stellarparams}. The companion Gl\,725B has a mass of 0.25$\pm$0.02\,M$_{\odot}$ and a radius of 0.280$\pm$0.005\,R$_{\odot}$\citep{Cristofari2022b}. Using Gaia astrometry, \citet{Kervella2019} ruled out the presence of a third component in the system. A more detailed analysis of Gl\,725B will be presented in Ould-Elhkim et al. (in prep). 

\subsection{Atmospheric parameters: $T_{\rm eff}$, \logg, and {\rm[Fe/H]}}

Stellar atmospheric parameters derived using the SPIRou near-infrared spectra of Gl\,725A are presented in \citealt{Cristofari2022}, on which the effective temperature ($T_{\rm eff}$), surface gravity (\logg), overall metallicity ([M/H]), and the alpha-enhancement parameter ([$\alpha$/Fe]) were computed. The results are listed in Table \ref{tab:stellarparams}.

We computed the $T_{\rm eff}$, \logg, and metallicity ([Fe/H]) using the SOPHIE optical spectra. These spectra are first shifted by the stellar RV and BERV, and corrected of the background pollution due to the calibration lamp \citep{Hobson2019}. Then, a high-S/N template was built by co-adding the corrected spectrum. We used this template as the input for the \texttt{ODUSSEAS}\footnote{\url{https://github.com/AlexandrosAntoniadis/ODUSSEAS}} code \citep{Antoniadis2020}, which computes the $T_{\rm eff}$ and [Fe/H] of M dwarf stars. 

The \texttt{ODUSSEAS} reference data set consists of spectra taken from the HARPS M dwarf sample, whose $T_{\rm eff}$ and [Fe/H] are well characterized. In the updated version of \texttt{ODUSSEAS}, the $T_{\rm eff}$ for this sample is computed through interferometry \citep{Rabus2019, Khata2021}, while [Fe/H] is derived by following the method by \citet{Neves2012}, which uses parallaxes. The code measures the pseudo-equivalent width (pEW) of the absorption lines in the input spectra and compares them with with the reference HARPS sample. Since HARPS has a resolution of 115\,000, the HARPS reference spectra are convolved from their original resolution to the SOPHIE resolution of 75\,000. The results obtained with \texttt{ODUSSEAS} for \starA are [Fe/H]~$ = -0.19 \pm 0.13$ dex and $T_{\rm eff}=3433 \pm 68$ K. In particular, the $T_{\rm eff}$ derived independently using SOPHIE and SPIRou spectra agree within the uncertainties. For comparison, the SPIRou $T_{\rm eff}$ is derived by comparing the observed spectra with grids of synthetic atmospheric models (see \citealt{Cristofari2022}).

Using the atmospheric parameters obtained with \texttt{ODUSSEAS}, $T_{\rm eff}$ and [Fe/H], along the Gaia parallax \citep{GaiaDR3}, we derived the \logg\, based on the SOPHIE spectra following the method by \citet{Sousa2021}. We find \logg $= 4.94 \pm 0.33$, which is in agreement with the value derived from SPIRou spectra by \citet{Cristofari2022b}.

\subsection{Mass and radius}

The stellar mass and radius were computed applying the empirical relationships for M dwarfs from \citet{Mann2015,Mann2019} using the [Fe/H], distance, and 2MASS \textit{K} magnitude (see Table \ref{tab:stellarparams}) as inputs. We obtained $R_{\star} = 0.351 \pm 0.003\,R_{\odot}$ and $M_{\star} = 0.333\pm0.008\,M_{\odot}$. Those values are in 1-$\sigma$ agreement with the results from \citet{Cristofari2022b}.

From the derived mass and radius, we estimate a \logg~$=4.87\pm0.01$. However, we treat this result with caution due to its small uncertainties and may indicate a underestimation of the mass and radius uncertainties. This value of \logg\, is somehow in between the derivations made with SOPHIE and SPIRou spectra independently and within 1-$\sigma$ to the SOPHIE \logg.

\subsection{Galactic population}

We used the public available code \texttt{GalVel\_Pop}\footnote{\url{https://github.com/vadibekyan/GalVel_Pop}}\citep{Adibekyan2012} to identify the galactic population of \starA. The code follows \citet{Reddy2006}, on which the position, proper motion, and parallax from Gaia DR3 \citep{GaiaDR3} are used to compute the galactic velocities and the probability to belong to the thin disk, thick disk, or halo. For \starA, the galactic velocities are (U, V, W) = (13.4, 0.5, 32.5) kms$^{-1}$, which indicates a $96.1\%$ probability of being part of the thin disk, and $3.9\%$ of the thick disk. The mean metallicity of stars in the thin disk is \hbox{<[Fe/H]> = $-0.09\pm0.19$} \citep{AllendePrieto2004}. The derived metallicity of \starA is within 1$\sigma$ of the mean value.

\subsection{Age}

Obtaining the age of M dwarfs is a challenging task and is usually poorly constrained. We used the Bayesian tool \texttt{PARAM} \citep{daSilva2006,Rodrigues2014,Rodrigues2017} to obtain an estimation of the stellar age. \texttt{PARAM} takes models from a grid of evolutionary tracks from the PARSEC code \citep{parsec} and matches them with stellar observables computed previously (i.e., [Fe/H], \logg), Gaia DR3 parallax, and \textit{V}, \textit{B}, \textit{J}, \textit{H}, \textit{K} magnitudes (see Table \ref{tab:stellarparams}). We obtained an age of $7.3^{+4.4}_{-4.7}$\,Gyr. Independently, using the gyrochronological relationship from \citet{Engle2023} calibrated for dwarfs with M2.5-M6.5 spectral types with rotation periods longer than 24\,days, we obtained an estimation of $6.2\pm2.3$\,Gyr. In previous works, the age of \starA was estimated at $14.5\pm5.4$\,Gyr by \citet{Fouque2023} whose age estimations are based on the methodology of \citet{Gaidos2023}. The upper part of this age range is limited by the age of the galactic thin disk (7.4-8.2 Gyr, \citealt{Kilic2017}).

The stars that belong to the thin disk, such as \starA, have ages in a wide range of values. However, most of them are less than 5\,Gyr with upper limits of 14\,Gyr \citep{Reddy2006,Haywood2008,Holmberg2009}. Our age estimations are in agreement with a star younger than 14\,Gyr and confirm an overestimation of the age derived by \citet{Fouque2023}.

Low-mass stars spin down with age as they lose angular momentum \citep{Barnes2007}. A rotation period of longer than 70\,days, as we have measured for \starA, is expected for stars older than 5\,Gyr \citep{Newton2016,Medina2022}. 

\subsection{Orbital evolution of binary system}\label{sec:binary}

Gl\,725 is a well-known binary system with observational data spanning from 1850 to 2020 gathered in the Washington Double Star Catalog (WDS). These archival observations consist of 714 measurements of the position angles (PA) and angular separation ($\rho$). We used the Bayesian parameter estimation tool \texttt{orbitize!}\footnote{\url{https://orbitize.readthedocs.io/en/latest/index.html}}\citep{Blunt2020} to obtain updated orbital parameters (semi-major axis $a_{\rm bin}$, eccentricity $e_{\rm bin}$, inclination $i_{\rm bin}$, $\omega$, $\Omega$, and phase $\tau$) and study the evolution of the components' position in the sky, as this can be critical to obtain high-precision individual photometric observations. The available PA and $\rho$ values were used as input for \texttt{orbitize!}. Observations taken previous than 2021 do not have uncertainties, hence we adopted values of $\pm50$\,deg for the PA and $\pm2.5$\,mas for $\rho$. 

To initialize the model, we adopted Gaussian priors in the parallax using the value from Table \ref{tab:stellarparams}, and in the total mass of the system, with a value of $M_{\rm T} = 0.58\pm0.02\,M_{\odot}$, computed using the mass of the primary component from Table \ref{tab:stellarparams} and the mass of the secondary component from \citet{Cristofari2022}. The priors of the orbital parameters are listed in Table \ref{table:binary_results}. In particular, to prevent the degeneracy in $\omega$ and $\Omega$, we adopted uniform priors for these parameters. The code uses a parallel-tempered Affine-invariant sampler \citep{emcee} to obtain the posterior distribution of the Markov-Chain Monte Carlo (MCMC) routine. 

\begin{figure*}
  \centering
   \includegraphics[width=0.9\linewidth]{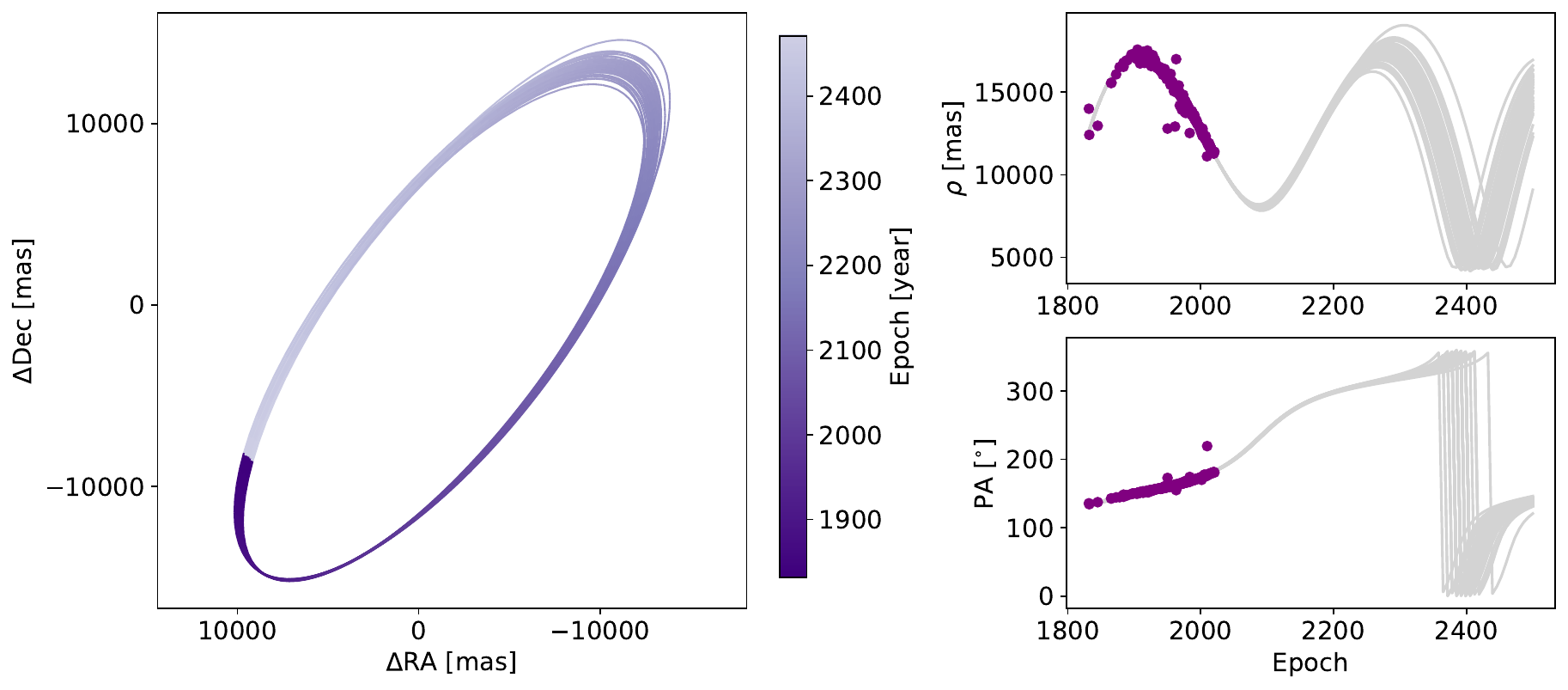}
  \caption{Orbit model of the Gl~725 binary system obtained with \texttt{orbitize!}. The figures show the results of 50 models randomly sampled from the posterior distributions obtained from the MCMC routine described in Sect. 4.6. \textit{Left: } Orbit of Gl\,725B with respect to Gl\,725A. \textit{Top right: } Model of the angular separation $\rho$ for one orbit. \textit{Bottom right: } Model of the position angle PA for one orbit.}\label{fig:orbit_binary}%
   \end{figure*}

We run the MCMC with 20 temperatures, 1000 walkers per temperature, and 100\,000 steps. We obtained a semi-major axis of $a_{\rm bin} = 63\pm1$\,AU, eccentricity of $e_{\rm bin} = 0.29\pm0.01$, and inclination of $i_{\rm bin} = 69.8\pm0.4$\,deg. The posterior distributions for all the model parameters are shown in Fig. \ref{fig:corner_binary} and Table \ref{table:binary_results} lists the final results. From them, we derived an orbital period of the binary system of $P_{\rm bin} = 871\pm108$ years. Figure \ref{fig:orbit_binary} shows the results of 50 models randomly sampled from the posterior predictive distributions. Our results are better constrained than the previous determination of the orbital parameters of this system by \citet{Izmailov2019}. In particular, we obtained a smaller eccentricity and larger semi-major axis, therefore longer orbital period. The main reason of the improvement in the model could be the different methods applied. While \citet{Izmailov2019} determined the orbital parameters through the Thiele-Innes method, \texttt{orbitize!} solves the Kepler equations of motion and also includes a MCMC procedure to sample the posterior distributions.

\begin{table}[ht]
\caption{Results of the binary orbit model of the binary system Gl\,725.}         
\label{table:binary_results}      
\centering
\footnotesize
\setlength{\tabcolsep}{2pt}
\renewcommand{\arraystretch}{1.3}
\begin{tabular}{lcc}     
\hline\hline
Parameter & Prior & Posterior \\ 
\hline
Semi-major axis $a_{\rm bin}$ [AU]  & $\rm \mathcal{U}(0,1000)$ & $63\pm1$\\
Eccentricity $e_{\rm bin}$ & $\rm \mathcal{U}(0,1)$ & $0.29\pm0.01$ \\
Inclination $i_{\rm bin}$ [deg] & $\rm \mathcal{U}(0,180)$ & $69.8\pm0.4$ \\
Argument of periastron $\omega$ [deg] &  $\rm \mathcal{U}(0,180)$ & $272.2^{+2.9}_{-3.0}$ \\
Longitude of ascending node $\Omega$ [deg] & $\rm \mathcal{U}(0,180)$ & $143.0\pm0.3$ \\
Phase $\tau$ & $\rm \mathcal{U}(0,1)$ & $0.62\pm0.01$ \\
Orbital period $P_{\rm bin}$ [years] & derived &  $871\pm108$\\
\hline
\end{tabular}
\tablefoot{$ \rm \mathcal{U}(a,b)$ describes a uniform prior, with $a$ and $b$ being the minimum and maximum limits, respectively.}
\end{table}


\section{Stellar activity analysis}\label{sec:activity}
The SOPHIE and SPIRou RVs show, independently, a significant periodicity at 11.2 days (see Sect. \ref{sec:obs}) with a $\log$(FAP)=-2.2 and \hbox{-1.4}, respectively. Their General Lomb-Scargle periodograms are shown in Fig. \ref{fig:periodograms}. In this section we analyse the stellar activity of Gl\,725 A to rule it out as the origin the signal at 11.2 days, which is analysed in details in Sect. \ref{sec:planet}.

We obtained stellar activity indicators from SOPHIE and SPIRou. On the optical side, we computed standard activity proxies from the CCF and spectral lines, such as: BIS, CCF FWHM, CCF contrast, H$\alpha$ and Ca \Romannum{2} H$\&$K lines which are used to compute the Mount Wilson S index \citep{Wilson1968,Baliunas1995}. Both activity indicators, the H$\alpha$ line and the S index, are computed following \citet{Boisse2009}. In the nIR domain, we obtained the FWHM, chromatic velocity slope (CVS), and longitudinal magnetic field (B$_\ell$). The FWHM is the second derivative of the spectral profile in units of FWHM, as defined in \citet{Artigau2022}, while the CVS is defined as the RV gradient as a function of wavelength \citep{Zechmeister2018}. Both activity indicators are computed in the LBL reduction. Tables \ref{table:SOPHIEdata} and \ref{table:SPIRoudata} of the Appendix list the activity indicators values for all the observations of SOPHIE and SPIRou, respectively. 

\subsection{Correlation with the RVs}

We computed the Pearson's coefficient between the RVs and the activity indicators to verify if the RV variability is linked to stellar activity \citep[e.g.,][]{Queloz2001,Boisse2011}. We found that none of the optical and nIR activity proxies are significantly correlated or anti-correlated to the RVs. In the optical part, all the activity indicators have Pearson's coefficient between -0.3 and 0.3. The nIR activity indicators are not correlated with the RVs with Pearson's coefficients between -0.1 and 0. These results are not surprising due to the slow rotation of the star that hampers the detection of spectral lines deformation \citep[e.g.,][]{Boisse2011}. However, it is also important to remark that the lack of correlation or anti-correlation can be also due to phase shifts between the RVs and the activity indicators \citep[e.g.,][]{Bonfils2007,Aigrain2012,CollierCameron2019}. We can discard this possibility as the activity indicator periodograms do not show the same periodicities as one would expect for signals with a time shift.

\subsection{Rotation period}

The stellar rotation period has been previously measured by \citet{Fouque2023} and \citet{Donati2023}. Both works applied a GPR framework with a quasi-periodic kernel in the SPIRou B$_\ell$ time series. However, they applied different data reduction procedures. In these works, the temporal variability of the B$_\ell$ has proved successful in measuring the rotation period of M dwarfs, even for quiet, slow-rotator stars. The reported rotation periods are $P_{\rm rot} = 103.5^{+4.6}_{-5.1}$\,d \citep{Fouque2023} and $P_{\rm rot} = 103.1\pm6.1$\,d \citep{Donati2023}, which are in 1-$\sigma$ agreement and confirm the slow-rotator nature of the target. 

The main differences between the two derivations of the rotation period are first, the data reduction software used, and second, the flexibility given in the GP fitting. While in \citet{Fouque2023} the evolution timescale and the smoothing factor are fixed, these hyper-parameters are left free in \citet{Donati2023}. In this work, we adopt the rotation period reported by \citet{Donati2023} since the final RMS of their model is smaller by a factor of two.

We studied the temporal variability of the activity indicators to see if they are modulated by the stellar rotation. The General Lomb-Scargle periodograms of all the activity indicators considered in this work are shown in Fig. \ref{fig:periodograms}, where a yellow dashed line marks the stellar rotation period. None of the activity indicators analysed show significant periodicities (FAP < 1$\%$) at the period of the stellar rotation, except for B$_\ell$. This is not surprising since the B$_\ell$ was previously used to determine the stellar rotation period of this star \citep{Fouque2023,Donati2023}. The lack of significant signals in the activity indicators periodograms may be related to the low activity of the star.

\section{Characterization of Gl 725A b}\label{sec:planet}


\subsection{Confirmation of the planet signal at 11.2 days}\label{sec:detection}
Since the period of the binary system is of the order of ~870 yr (see Sect. 3.5), the effect of the stellar companion Gl\,725 B in the SOPHIE and SPIRou RVs was corrected by a simple linear fit. In particular for the SPIRou RVs, the linear trend is corrected within the \texttt{Wapiti} correction. The contribution of the binary companion in the RVs is of 5.9 m/s/yr. The mean RV of each data set is subtracted to take into account the offsets of the instruments.

We analyse the periodicities of the SOPHIE and SPIRou RVs by computing their Generalised Lomb-Scargle periodograms, which are shown in Fig. \ref{fig:periodograms}. Both data sets independently show a power excess at the period of 11.2 d. This period gets a higher power when the SOPHIE and SPIRou RVs are combined into one data set with a $\log{\rm( FAP)}=-5.4$, corresponding to the period with the highest power in the periodogram. The stellar rotation signal may be seen in the periodograms of the SPIRou RVs at a period of 111.9 d, which is somehow longer than the expected stellar rotation period but still within uncertainties. However, we can not completely rule out additional influence of residuals from the \texttt{Wapiti} correction (see Sect. \ref{wapiti}). 

\begin{figure}
  \centering
   \includegraphics[width=\linewidth]{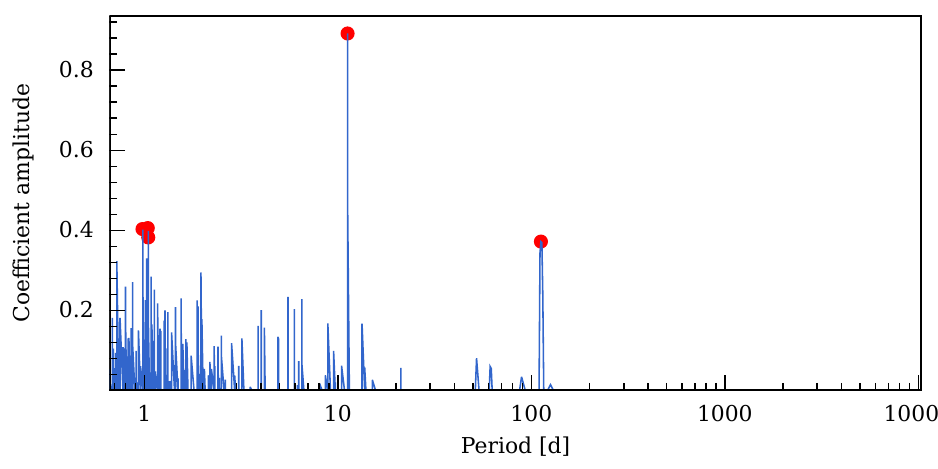}
  \caption{$l_1$ periodogram of the SOPHIE and SPIRou RVs. The five most significant periods are highlighted with a red circle. The signal with the highest amplitude is located at 11.2 d with a $\log_{10}(\rm{FAP})=-5.9$. Around 1 day we see three peaks which origin is the data sampling. The longest significant periodicity is located at 113 days.}\label{fig:l1periodogram}
   \end{figure}

We also computed the $l_1$ periodogram from \citet{Hara2017}, which works in a similar way as the Lomb-Scargle periodogram in terms of searching for periodicities in time series. However, the $l_1$ periodogram shows less number of peaks coming from aliasing of the signals, resulting in a cleaner periodogram. The $l_1$ periodogram is shown in Fig. \ref{fig:l1periodogram} and it confirms that the signal with a period of 11.2 d is the most significant in the data set. In this figure it is also seen some peaks at 1 day, which correspond to the 1-day periodicity due to sampling, and an isolated signal at 113 days, also spotted in the Lomb-Scargle periodogram.

\begin{figure}
  \centering
  \includegraphics[width=\linewidth]{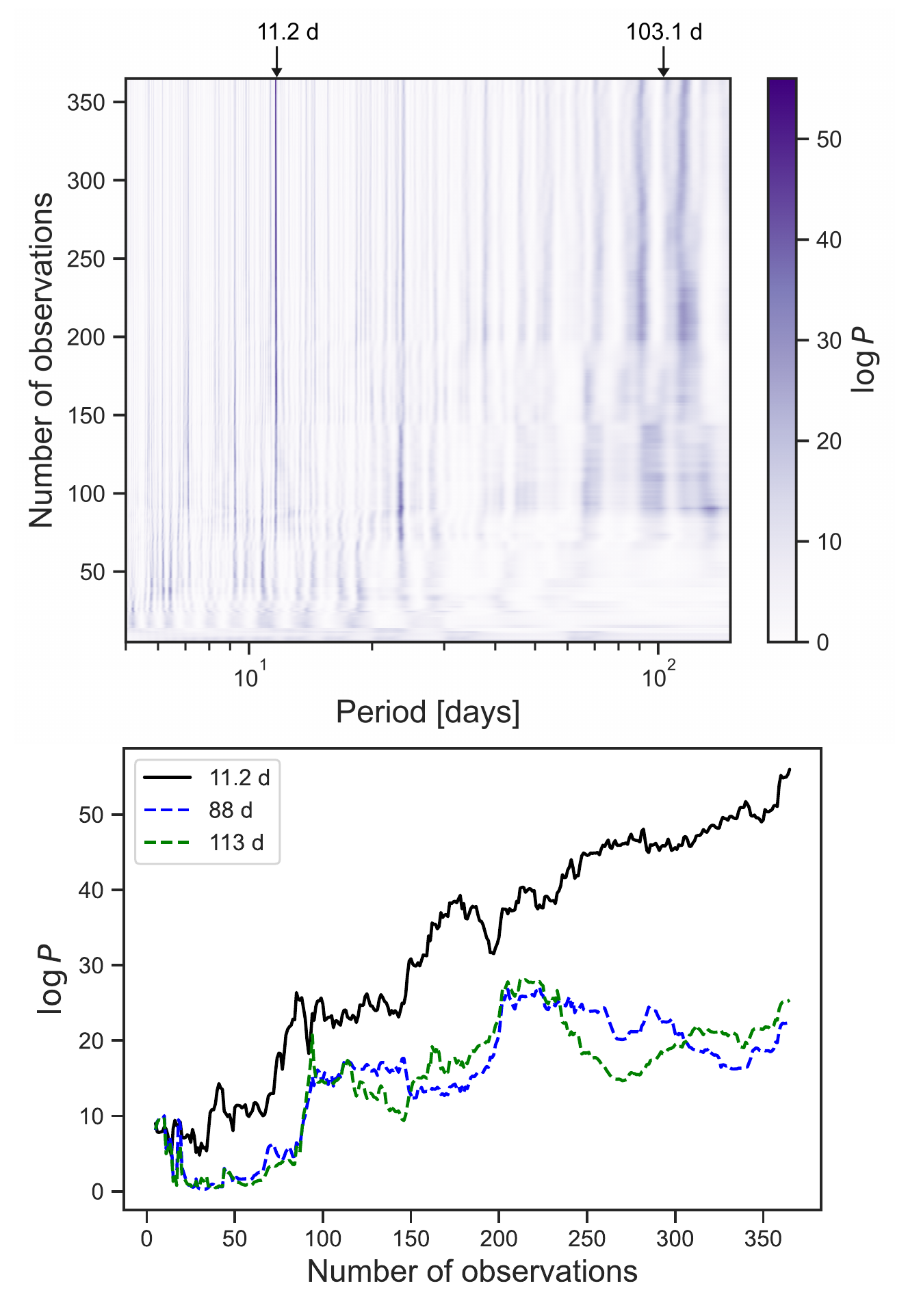}
  \caption{Results of the BGLS periodogram analysis. \textit{Top:} stacked BGLS periodograms of the complete RV data set. The colour bar indicates the log-likelihood (log$P$) level. The black arrows mark the planetary signal at 11.2 days and the stellar rotation period at 103.1 days. \textit{Bottom:} Evolution of the log-likelihood a function of the number of observations for the three most significant signals in the BLS periodogram: 11.2, 88, and 113 days.}\label{fig:logP_RV}
   \end{figure}

We applied the Stacked Bayesian general Lomb-Scargle (BGLS) periodogram \citep{Mortier2017} to study the coherence and significance of the signals spotted at 11.2 and 111.9 days. The log-likelihood of the signal at 11.2 days increases consistently as more observations are added in the stacking (see Fig. \ref{fig:logP_RV}), as it is expected for signals of planetary nature. We identified a second group of strong signals surrounding the rotation period, in particular at 88 and 113 d as seen in the BGLS periodogram. These signals, on the contrary, do not show a consistent increasing log-likelihood and show a complex temporal evolution (see bottom panel of Fig. \ref{fig:logP_RV}). This is an indication that are related to activity or systematics.

Furthermore, the signal at 11.2 days is not found in any of the activity indicators periodograms as seen in Fig. \ref{fig:periodograms}, confirming its planetary nature, while signals with periods longer than \hbox{100 days} are seen in the periodograms of the B$_\ell$, H$\alpha$, and SPIRou FWHM.

To confirm the consistency between instruments of the planetary signal at 11.2 days, we fit independently a Keplerian model in the SOPHIE (Planet$_{b,\rm{SOPHIE}}$) and SPIRou (Planet$_{b,\rm{SPIRou}}$) RVs. We used the \texttt{RadVel}\footnote{\url{https://radvel.readthedocs.io/en/latest/}}\citep{Fulton2018} Python package to model the planetary signal detected at 11.2 days with a Keplerian model ($P_{b}$, $T\rm{conj}_{b}$, $e_{b}$, $\omega_{b}$, $K_{b}$) alongside the mean RV ($\gamma_{\rm SOPHIE}$, $\gamma_{\rm SPIRou}$) and a white noise term ($\sigma_{\rm SOPHIE}$, $\sigma_{\rm SPIRou}$). The posterior distributions of the parameters were obtained using a Markov-chain Monte Carlo (MCMC) routine using the package \texttt{emcee}\citep{emcee}, set up with 50 walkers and 10\,000 steps after a burn-in phase, which runs until the Gelman-Rubin statistic is less than 1.03 so the chains are marginally mixed. After the burn-in phase, the final convergence of the chains is reached when the Gelman-Rubin statistic is less than 1.01 \citep{Gelman2014}, suggesting that the chains are well-mixed. The priors applied in the parameters are listed in Table \ref{table:results}.

The results of the Keplerian models applied independently in the SOPHIE and SPIRou RVs are listed in the first two columns of Table \ref{table:results}. Both models are consistent within uncertainties between each other, reassuring a robust detection of the planetary signal at 11.2 days. Having detected a consistent planetary signal in both instruments with different wavelengths discards a stellar activity origin as the signal is achromatic. Both instruments have detected a planet with a small semi-amplitude between 1.6 and 1.7 m/s in a short-period orbit of 11.2 days. 

\subsection{Planet model selection}\label{sec:keplerian}


After confirming the signal independently in both instruments, we applied a joint Keplerian model (Planet$_{b,\rm{SOPHIE+SPIRou}}$) in the whole dataset of SOPHIE and SPIRou RVs. We followed the same procedure using \texttt{RadVel} as in Sec. \ref{sec:detection} and applied the same priors listed in Table \ref{table:results}. In particular for the eccentricity, we tested leaving it as a free parameter and fixed in $e=0$. We found the best model by comparing their Bayesian Information Criteria (BIC), following \citealt{BIC}. The model with fixed zero eccentricity is preferred with a $\Delta\rm{BIC}=8.8$ (see Table \ref{table:results}).



   \begin{figure*}
  \centering
   \includegraphics[width=0.96\textwidth]{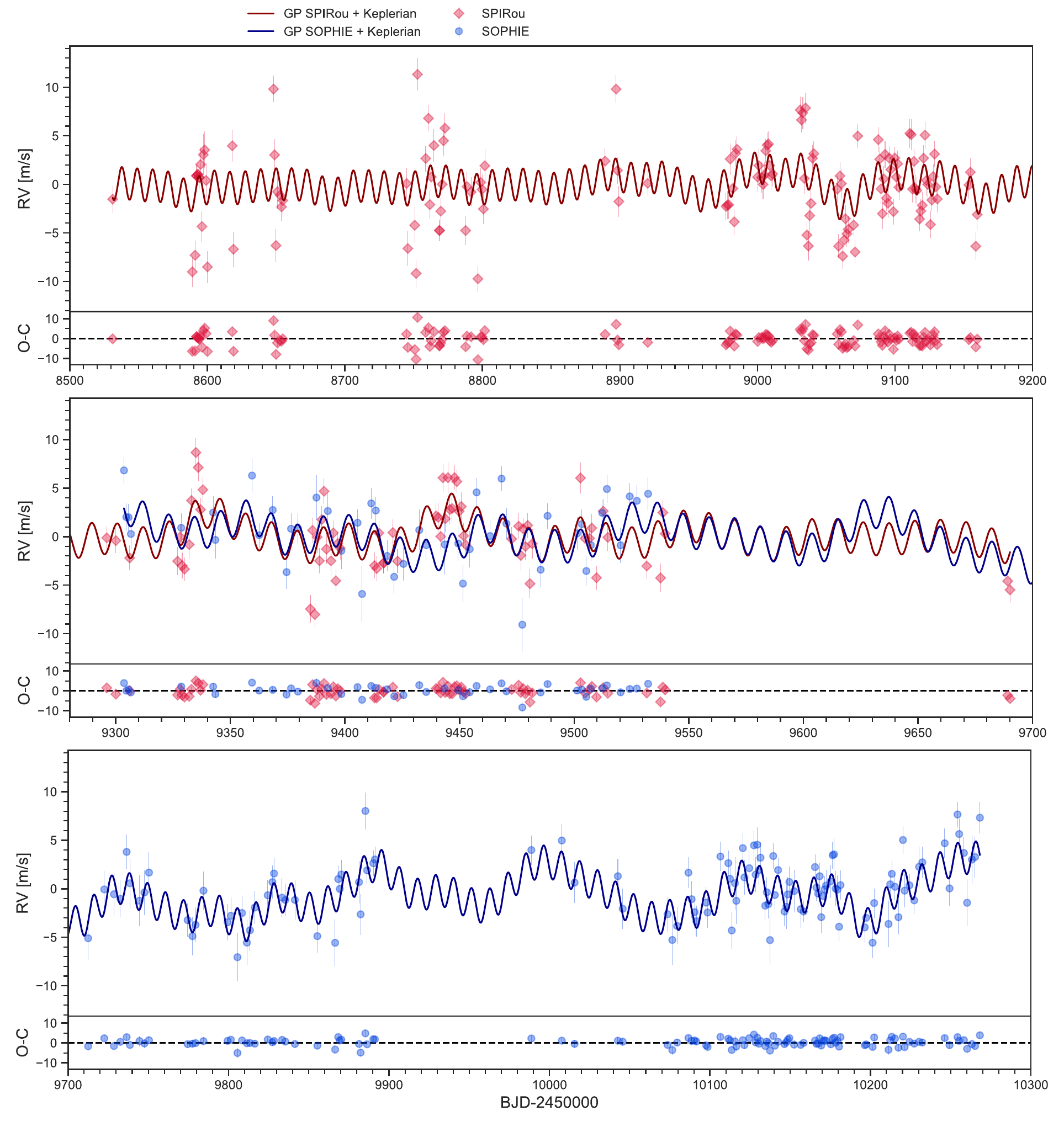}
  \caption{RVs time series of SOPHIE in blue circles, and SPIRou in red squares. The best-fit model including the Keplerian signal and the GP for each instrument are depicted as a dark blue line for SOPHIE and dark red line for SPIRou. The panels represent three consecutive observing seasons,  with their respective residuals shown underneath. The residuals of the final model yield RMS are 2.0\,m/s for SOPHIE and 3.2\,m/s for SPIRou.}\label{fig:rv_model}
   \end{figure*}

We increased the complexity of the single Keplerian model by simultaneously modelling the Keplerian with the stellar activity modulation in the RVs. We assumed that the stellar activity signal evolves quasi-periodic, as it has been shown in previous studies \citep[e.g.,][]{Haywood2014,Angus2018}, and took advantage of the flexibility of GPR to model this signal. The covariance matrix of the quasi-periodic kernel implemented in \texttt{RadVel} is defined as shown in Equation \ref{eq:kernel2}.

\begin{equation}\label{eq:kernel2}
   C_{i,j} = \eta_{1}^{2} \text{exp}\bigg[-\frac{(t_{i}-t_{j})^2}{\eta_{2}^2}- \frac{1}{\eta_{4}^2}\text{sin}^2\bigg(\frac{\pi(t_{i}-t_{j})}{\eta_{3}}\bigg)\bigg] + \sigma^2\delta_{i,j}
\end{equation}

\noindent where $t_i$ and $t_j$ are two observation dates, $\eta_{1}$ is the amplitude of the covariance, $\eta_{2}$ is the non-periodic characteristic length, $\eta_{3}$ is the period of the signal, $\eta_{4}$ is the periodic characteristic length or harmonic complexity, and $\sigma$ is the uncorrelated white noise also known as the jitter term. These hyper-parameter are linked to physical parameters of the stellar activity signal \citep[e.g.,][]{Nicholson2022,Stock2023}, such as the stellar rotation period ($\eta_{3}$) and the decay time of the active regions in the stellar surface ($\eta_{2}$).


We simultaneously fit one Keplerian model and two independent GPs, one for each instrument, in the whole data set. This is because while the stellar rotation period is consistent across wavelength, the amplitude of the activity modulation can change, especially in the case of M dwarfs observed with optical and near-infrared instruments \citep[e.g.,][]{CortesZuleta2023}. We applied wide uniform priors in the mean of the time series, amplitude $\eta_{1}$, harmonic complexity $\eta_{4}$, decay time $\eta_{2}$, and white noise $\sigma$. We defined a more constraining prior for $\eta_{3}$ as the rotation period was measured from the SPIRou spectropolarimetric data, in the form of a Gaussian distribution $\mathcal{N}(103.1,6.1)$. All the priors are listed in Table \ref{table:results}. For this model including GPs the eccentricity is fixed to zero, and we compared the models sharing GP hyper-parameters between the two instruments: one model with shared stellar rotation period $\eta_{3}$ ($\eta_{3, \rm SOPHIE}$=$\eta_{3, \rm SPIRou}$), and a second model with shared stellar rotation period $\eta_{3}$ and decay time $\eta_{2}$ ($\eta_{2,3, \rm SOPHIE}$=$\eta_{2,3, \rm SPIRou}$). This is motivated because the optical and near-infrared RVs show similar scatters which suggest that there is no evident chromaticity in the RVs. Therefore, active regions should modulate optical and near-infrared data in similar ways. 

\begin{figure}
  \centering
   \includegraphics[width=\linewidth]{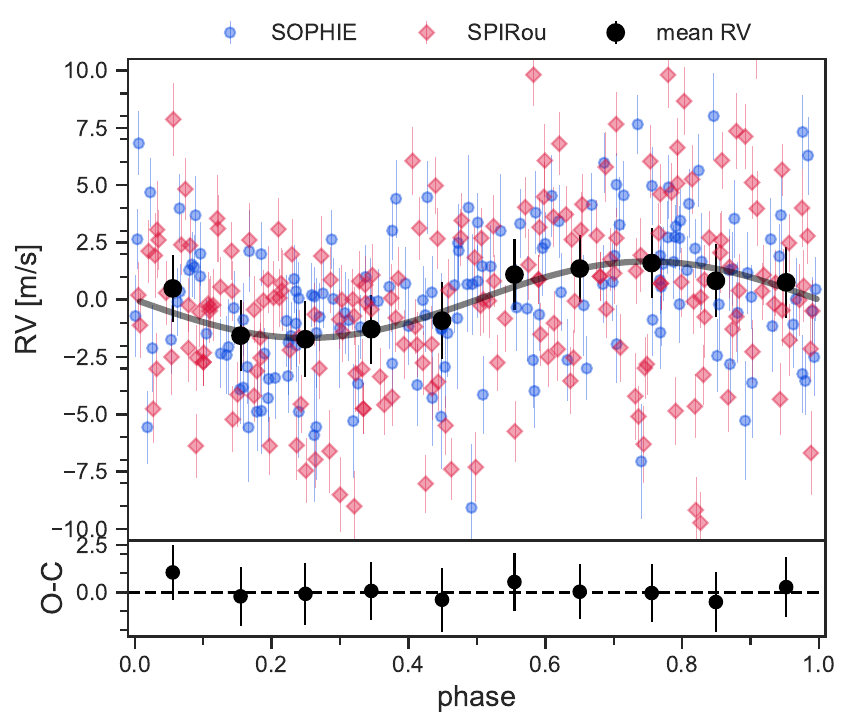}
  
  \caption{Phase-folded RVs of SOPHIE in blue circles, and SPIRou in red squares. In black are represented the mean RV for each phase bin. The black line indicates the best-fitted Keplerian model of a planet with an circular orbit of 11.2 days period and a semi-amplitude of $1.67\pm0.20$\,m/s.}\label{fig:phase_rv}
   \end{figure}

The model with $e=0$, and shared GP rotation period and decay time ($\eta_{2,3, \rm SOPHIE}$=$\eta_{2,3, \rm SPIRou}$) strongly preferred with a $\Delta\rm{BIC}=14.6$ when compared with the most simple Keplerian model without GPs. It is important to highlight that even though its BIC is very close to the model with just shared stellar rotation period $\eta_{3}$, the model planetary parameters are identical. Therefore, the adopted model for Gl\,725A b is a circular Keplerian and two GPs to account for the stellar variability with shared stellar rotation period $\eta_{3}$ and decay time $\eta_{2}$. The best-fit models for each instrument and their residuals are shown in Fig. \ref{fig:rv_model}, on which the scatter of the residuals is 2.7\,m/s. The final posterior distributions of the parameters are listed in Table \ref{table:results}, and their corner plot is displayed in Fig. \ref{fig:corner}. Figure \ref{fig:phase_rv} shows the phase-folded Keplerian model. For comparison, the model without a Keplerian (null hypothesis) has a $\rm{BIC}=5328.9$. This indicates a strong detection of the planet signal because all the models including a Keplerian have much smaller BIC values (see Table \ref{table:results}).

The adopted model describes a planet in a circular orbit with an orbital period of $11.2201\pm0.0051$ days and a RV semi-amplitude of $1.67\pm0.20$ m/s, indicating a significant detection. We derived a minimum planet mass of $M_{p}\sin{i}=2.78\pm0.35\,M_{\oplus}$, and a semi-major axis of $0.068\pm0.001$\,AU. All the planet and instrumental parameters of the preferred model are listed in the last column of Table \ref{table:results}. The amplitude of the GP in the SOPHIE RVs ($\eta_{1, \rm{SOPHIE}}$) is larger than that of SPIRou, which could indicate a greater activity contribution in the optical data. On the other hand, the white noise term decreases for both instruments when including the GPs to account for stellar activity. This effect is even more pronounced for SOPHIE, whose white noise term decreases from $\sigma_{\rm SOPHIE}=2.22\pm0.20$\,m/s in the single Keplerian model, to $1.09^{+0.31}_{-0.52}$\,m/s when including the GP.

Gl\,725A b is too close to its host star to be in the Habitable Zone, which is located between 0.10 and 0.26 AU following the definition of \citet{Kopparapu2013}. Assuming an albedo of 0, the derived equilibrium temperature of the planet is of $T_{\rm{eq}} = 377\pm3$ K. If we use the Earth's and Venus' albedos (0.3 and 0.7), the $T_{\rm{eq}}$ is $345\pm3$ K and $279\pm3$ K, respectively.

Binary companions can align protoplanetary disks with the orbital plane of the binary system constraining the orbital inclination of their planets \citep{Lai2014,Zanazzi2018}. Recent observational evidence has suggested that S-type planets in close binary systems ($a_{\rm bin}<100 \,\rm{AU}$) show some preference to be aligned with the orbital plane of the binary system \citep[e.g.,][]{Behmard2022,Christian2022,Lester2023}. \citealt{Christian2024} found that this alignment is more significant for small planets ($R_{p}<6R_{\oplus}$) in closer binary orbits ($a_{\rm bin}<700\,\rm{AU}$). From the analysis of Sect. \ref{sec:binary}, we found the inclination of the binary system to be $i_{\rm{bin}}=69.7\pm0.4^{\circ}$. Assuming this value is the orbital inclination of the planet, we can place a potential mass planet of $M_{p} = 3.0\pm0.4\,M_{\oplus}$. This value is in agreement with the derived minimum mass from the RV analysis.

\subsection{Detection limits for additional companions}

\begin{figure}
  \centering
  \includegraphics[width=\linewidth]{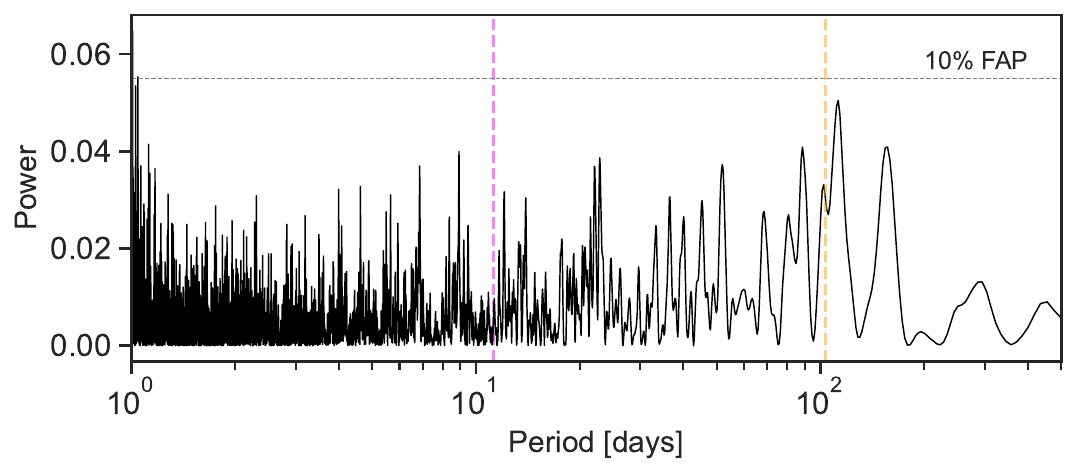}
  \caption{Periodogram of the RV residuals of the adopted model for Gl\,725A b which includes two GPs, one for each instrument, and one Keplerian. For more details see Sect. \ref{sec:detection}. The dashed horizontal line indicates 10$\%$ of FAP and the magenta and orange vertical lines indicate the period of Gl\,725A b and the stellar rotation, respectively.}\label{fig:RV_residuals}
   \end{figure}

The periodogram of the residuals of the two GPs and Keplerian model has no periodicities left with power below 10$\%$ FAP (see Fig. \ref{fig:RV_residuals}), except for the 1-day signal coming from the daily observations. This suggests that the signals from planet b and the stellar rotation and/or instrumental systematics are the only signals detectable in our data set. However, the instrumental sensitivity and the observing time span can hamper the detection of planets with lower masses or in longer orbital periods. To determine the detection limits in our RV data set we applied a simple injection-recovery test for companions with eccentricity equals zero in a activity-only RV data set, on which the signal of planet b was subtracted. 

We created 10\,000 sinusoidal models in a log-space grid of 200 periods between 1 and 800 days and 50 semi-amplitudes between 0.1 and 10 m/s. The upper limit in the period sampling is determined to ensure at least two orbits in the data set, as our total time span is $\sim1\,600$ days. The phase of each model is randomly sampled from an uniform distribution.

\begin{figure}
  \centering
  \includegraphics[width=\linewidth]{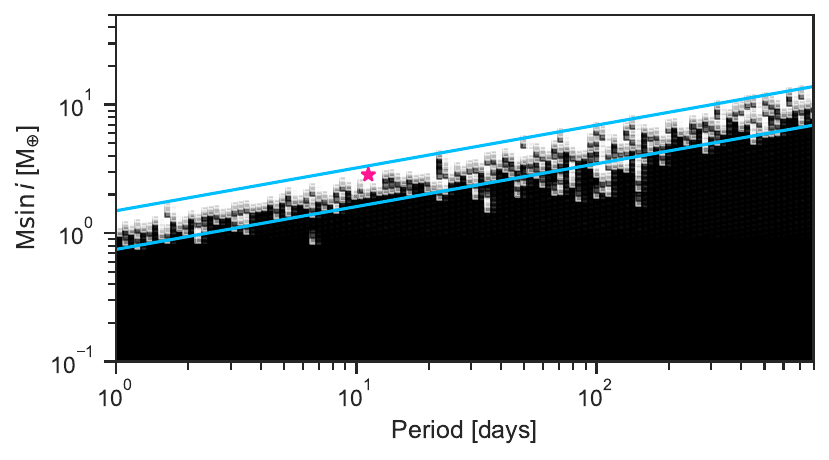}
  \caption{Results for the injection-recovery test in the RV residuals after subtracting the signal of Gl\,725A b. The black zone is the parameter space of undetectable signals using a GLS periodogram, i.e with a FAP below 1$\%$, whereas the white area is that for signals above 1$\%$. The magenta star depicts the position of Gl\,725A b, with a semi-amplitude of 1.67\,m/s and an orbital period of 11.2\,days. To help guiding the eye, the light blue lines indicate the semi-amplitudes of 2 and 1 m/s.}\label{fig:RV_injection}
   \end{figure}

For each model, we computed a GLS periodogram to measure the power of the injected signal. Signals with FAP lower than $1\%$ are considered as detectable. The results are shown in Fig. \ref{fig:RV_injection}, where in white is the parameter space of detectable signals. We notice from these results that most of the low amplitude signals (i.e., at the level of Gl\,725A b) could be recovered for short and long orbital periods. Signals with semi-amplitudes below 1 m/s are hardly detectable in our data set. Moreover, we can confidently discard the presence of planets with $M_{p}\sin{i}>10\,M_{\oplus}$ in the period range between 1 and 800 days orbiting Gl\,725 A.

\begin{landscape}
\begin{table}[ht]
\caption{Results of the models applied in the SOPHIE and SPIRou data, independently and jointly. Below each model's name is indicated if the eccentricity is fixed to $e=0$ or free, and which GP hyper-parameters are shared between the instruments.}         
\label{table:results}      
\centering
\footnotesize
\setlength{\tabcolsep}{2pt}
\renewcommand{\arraystretch}{1.5}
\begin{tabular}{ccc|c|c|cc|cc}     
\hline\hline
 & &  & Planet$_{b,\rm{SOPHIE}}$  & Planet$_{b,\rm{SPIRou}}$ & \multicolumn{2}{c|}{Planet$_{b,\rm{SOPHIE+SPIRou}}$} & \multicolumn{2}{c}{GP$_{\rm SOPHIE}$ + GP$_{\rm SPIRou}$ + Planet$_{b,\rm{SOPHIE+SPIRou}}$}  \\
Parameter & Units & Prior & $e$ free      & $e$ free     &   $e$ free          &    $e=0$             & $e=0$, $\eta_{3,\rm{SOPHIE}}$=$\eta_{3, \rm SPIRou}$ &  $e=0$, $\eta_{2,3, \rm SOPHIE}$=$\eta_{2,3, \rm SPIRou}$\\

\hline
\multicolumn{9}{l}{Orbital parameters}\\
\hline
$P_{b}$         & days   &  $\mathcal{U}(11,11.5)$   & $11.210\pm 0.012$       & $11.238^{+0.025}_{-0.018}$  &$11.2155^{+0.0065}_{-0.0052}$ & $11.2183\pm0.0060$ & $11.2202\pm0.0049$ & $11.2201\pm0.0051$\\
$T\rm{conj}_{b}$& BJD-2450000 & $\mathcal{U}(8591,8602)$ & $8598.4^{+1.5}_{-1.7}$ & $8596.3^{+1.5}_{-1.9}$      & $8597.8\pm1.0$               & $8597.0\pm0.6$               & $8596.7\pm0.5$ & $8597.7\pm0.5$   \\
$e_{b}$         &             & & $0.24^{+0.27}_{-0.17}$ & $0.23^{+0.23}_{-0.16}$      & $0.20^{+0.19}_{-0.14}$        & $\equiv0.0$                   &$\equiv0.0$ & $\equiv0.0$ \\
$\omega_{b}$    & rad     &    & $2.6^{+1.5}_{-1.1}$    & $-2.6^{+1.6}_{-2.4}$        & $2.9^{+1.20}_{-0.72}$         & $\equiv0.0$                   & $\equiv0.0$ & $\equiv0.0$\\
$K_{b}$         & m/s      & $\mathcal{J}(0.01,10)$  & $1.63^{+0.38}_{-0.35}$ & $1.69\pm 0.44$              & $1.69\pm 0.26$               & $1.61\pm 0.24$                & $1.67\pm0.20$ & $1.67\pm0.20$ \\
\hline
\multicolumn{9}{l}{Derived planet parameters}\\
\hline
$M_b\sin i$ & M$_{\oplus}$ & & $2.56\pm0.55$       & $2.68^{+0.67}_{-0.70}$          & $2.71\pm0.42$       & $2.69\pm 0.41$ & $2.79\pm0.35$ &$2.78\pm0.35$\\
$a_b$       & AU       &    & $0.068\pm0.001$ & $0.068\pm0.001$   & $0.068\pm0.001$ & $0.068\pm0.001$& $0.068\pm0.001$ & $0.068\pm0.001$\\
\hline
\multicolumn{9}{l}{Instrumental parameter posteriors}\\
\hline
$\gamma_{\rm SOPHIE}$ & m/s  & & $0.39^{+0.22}_{-0.23}$ &                         & $0.38\pm 0.22$         & $0.35\pm 0.22$ & $0.62^{+0.83}_{-0.73}$ & $0.61^{+0.90}_{-0.75}$\\
$\sigma_{\rm SOPHIE}$ & m/s & $\mathcal{J}(0.01,5)$ &$2.23\pm0.20$  &      & $2.2\pm0.20$  & $2.22\pm0.20$  & $1.08^{+0.31}_{-0.55}$ & $1.09^{+0.31}_{-0.52}$\\
$\gamma_{\rm SPIRou}$ & m/s &  &                       & $-0.02^{+0.25}_{-0.26}$ & $-0.04^{+0.26}_{-0.25}$& $-0.04\pm 0.26$ & $-0.35\pm0.61$ &$-0.26\pm0.47$\\
$\sigma_{\rm SPIRou}$ & m/s & $\mathcal{J}(0.01,5)$  &                       & $3.36\pm0.20$   & $3.35\pm0.20$ & $3.36\pm0.21$  & $3.01\pm0.21$ &$3.01\pm0.22$\\
\hline
\multicolumn{9}{l}{GP Hyper-parameters posteriors}\\
\hline
$\eta_{1, \rm SOPHIE}$&      &  $\mathcal{J}(0.01,5)$&  & &   & &$2.46^{+0.59}_{-0.43}$ & $2.52^{+0.67}_{-0.47}$ \\
$\eta_{2, \rm SOPHIE}$& days & $\mathcal{U}(100,500)$& & &    & &$125^{+29}_{-18}$& $\eta_{2,\rm{SOPHIE},\rm{SPIRou}}$\\
$\eta_{4, \rm SOPHIE}$&      & $\mathcal{U}(0.1,1.0)$ && &    & &$0.44^{+0.30}_{-0.19}$ & $0.45^{+0.31}_{-0.20}$\\
$\eta_{1, \rm SPIRou}$&      &$\mathcal{J}(0.01,5)$ & & &    &&$1.53^{+0.51}_{-0.40}$ & $1.43\pm0.40$\\
$\eta_{2, \rm SPIRou}$& days & $\mathcal{U}(100,500)$ & & &    & &$351^{+99}_{-130}$ & $\eta_{2,\rm{SOPHIE},\rm{SPIRou}}$\\
$\eta_{4, \rm SPIRou}$&      & $\mathcal{U}(0.1,1.0)$&  & &    && $0.32^{+0.22}_{-0.14}$ & $0.30^{+0.21}_{-0.14}$\\
$\eta_{2,\rm{SOPHIE,SPIRou}}$ & days & $\mathcal{N}(100,500)$ &  & &    & & &  $134^{+35}_{-23}$ \\
$\eta_{3, \rm{SOPHIE,SPIRou}}$& days & $\mathcal{N}(103.1,6.1)$ & & &    & &$105.0\pm2.6$ &  $104.8\pm3.2$  \\

\hline
\multicolumn{9}{l}{Model diagnostics}\\
\hline
RMS                & m/s & & 2.8    & 3.6    & 3.3    & 3.3      & 2.7 & 2.7\\
$\ln{\mathcal{L}}$ &     & & -413.3 & -546.5 & -957.6 & -959.1   & -945.9 & -941.9 \\
BIC                &     & &   &  & 1942.3 & 1933.5 & 1920.6 & 1918.9\\
$\Delta$BIC &  & &   &  & -8.8 & 0 & 13.1 & 14.6 \\
\hline
\end{tabular}
\tablefoot{The symbol $ \rm \mathcal{U}(a,b)$ describes a uniform prior, with $a$ and $b$ being the minimum and maximum limits, respectively. The symbol $\rm \mathcal{J}(a,b)$ describes a Jeffrey's prior, with $a$ and $b$ being the minimum and maximum limits, respectively. The symbol $\rm \mathcal{N}(a,b)$ describes a normal distribution centred in $a$ and standard deviation $b$.}
\end{table}
\end{landscape}

\section{Transit search in TESS data}\label{sec:tess}

\subsection{Predicted planetary radius}\label{sec:radius}

\begin{figure}
  \centering
  \includegraphics[width=\linewidth]{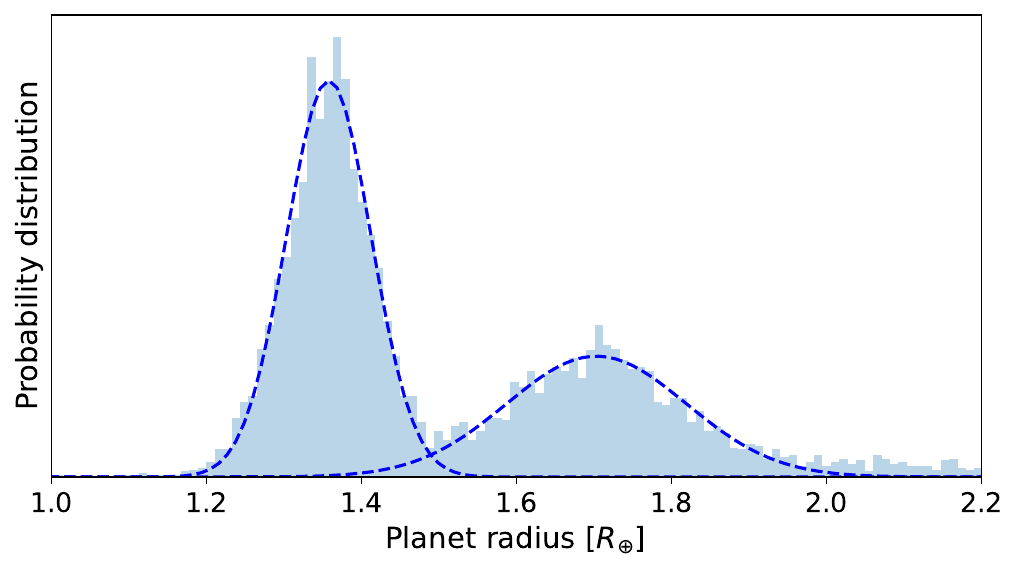}
  \caption{Probability density distribution of the predicted planet radius of Gl\,725A b obtained using \texttt{spright}, for a minimum mass of M$_{p}\sin{i}=2.78\pm0.35\, M_{\oplus}$. The bimodal distributions are described by two Gaussians depicted with dashed blue lines. 
  }\label{fig:spright_radius}
   \end{figure}

Mass-radius relationships are useful tools to forecast missing planet parameters based on known demographics. We used the tool \texttt{spright}\footnote{\url{https://github.com/hpparvi/spright}}\citep{Parviainen2024} which implements radius-density-mass relationship based on Bayesian inference. In particular, we applied a specific model for exoplanets around M dwarfs contained in \texttt{spright}, which is based on the Small Transiting Planets around M dwarfs (STPM) catalog by \hbox{\citet{Luque2024}}. For a planetary mass equals to the derived minimum mass of Gl\,725A b, the posterior distribution of the predicted radius is bimodal with peaks at $1.35\pm0.06\, R_{\oplus}$ and $1.71\pm0.11\, R_{\oplus}$, being the former radius the one with the highest probability, as seen in Fig. \ref{fig:spright_radius}. 

We obtained similar results with the \texttt{forecaster} tool \citep{Chen2017} which is based on a probabilistic model of the mass-radius relationship. Unlike \texttt{spright}, it does not differentiate on planets orbiting different types of stars. Using the \texttt{forecaster} model, the minimum mass of Gl\,725A b predicts a radius of $1.47^{+0.64}_{-0.37}\,R_{\oplus}$. 

The results obtained from \texttt{spright} and \texttt{forecaster} for the planet minimum mass, place Gl\,725A b most likely in the super-Earth regime (1.2 < $R_{p}$/$R_{\oplus}$ < 2.0), with a high probability of being $\sim1.4\,R_{\oplus}$. A planet in the sub-Neptune regime cannot be completely discarded from the predicted planet radius, however, the mass estimation from the binary orbit alignment allow us to put strong constraints in the radius of the planet.

\subsection{Transit detectability in the TESS light curves}

   \begin{figure}
  \centering
   \includegraphics[width=\linewidth]{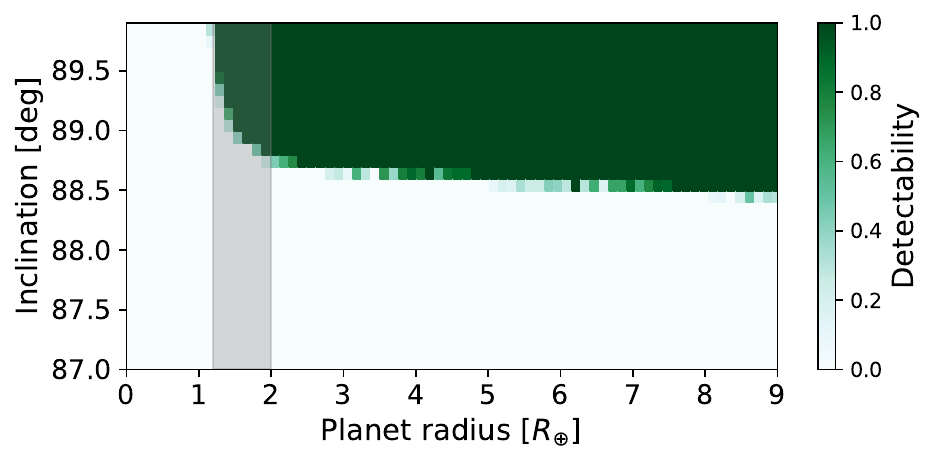}
   \caption{Detectability of transit events of the Gl\,725A b in the TESS light curves of Gl725A+B, as a function of the orbital inclination and the planet radius. The color bar indicates the probability of detection as described in Sect. 6. The grey area depicts the predicted planet radius range for the minimum mass of the planet.}\label{fig:detectability}
   \end{figure}

We used the transit least squares algorithm (TLS, \citealt{Hippke2019b}) to search for transit events in the blended light curves of Gl 725 A and B from TESS (see Sect. \ref{sec:obs}). No transits were identified by the TLS including periodic events at the orbital period of Gl\,725 A b. Given the long time span of the TESS data consisting of 27 Sectors, a periodic event of 11.2 days could be easy to identify. However, the derived planetary parameters from the RVs suggest a transit probability of only 1.7$\%$.

To assess the transit detection sensitivity, we conducted an injection-recovery test in our TESS data, for a range of orbital inclination and planet radius. Since the light curves are blended, the true planet radius detectable needs to be scaled by the dilution factor, which has a value of 1.26. We performed 50\,000 simulations of transit events generated by the \texttt{batman} Python package \citep{batman}. For each run, the orbital inclination and planet radius are drawn randomly from uniform distributions between 85$^{\circ}$ and 90$^{\circ}$ for the inclination, and 0.1 to 10\,$R_\oplus$ for the radius. Then, we applied the box least squares (BLS, \citealt{Kovacs2002}) periodogram\footnote{\url{https://docs.astropy.org/en/stable/timeseries/bls.html}} from Astropy to measure the power of the signal at the expected orbital period of 11.2 days.

We binned the orbital inclination and planet radius in bins of 0.1 units for each parameter and then computed the fraction of transit events that are detectable with the BLS periodogram (i.e., power > 2000) in each bin. The results are shown in Fig. \ref{fig:detectability}, where we found that planets with a radius greater than 1.2$R_\oplus$ and orbital inclinations between 88.5$^{\circ}$ and 90$^{\circ}$ can be detected in our TESS data. For orbital inclinations lower than 88.5$^{\circ}$, the planet is not transiting (up to a planet radius of 8$\,R_{\oplus}$). Importantly, from the results of Sect. \ref{sec:radius}, we expected a planet radius likely between 1.2 and 2$\,R_{\oplus}$, which is detectable in our available TESS data for planets with orbital inclinations close to 90$^{\circ}$. However, as we increase the planet radius, the orbital inclination range at which the planet is detectable also increase. For example, for a radius of 1.5$R_\oplus$, the transit is detectable up to 89$^{\circ}$, and for a radius of 2.0$R_\oplus$ is up to 88.7$^{\circ}$. From this test we can conclude that Gl\,725A b is likely a non-transiting planet.

\subsection{True Inclusion Probability of transits}

To further assess the existence of transits in the TESS data due to the discovered RV planet, we conducted a novel statistical analysis (Wilson et al., in review). In brief, we combined the TESS PSF-based PCA components produced following \citet{Wilson2022} with the instrumental co-trending basis vectors and quaternions to construct a linear noise model. This was fit simultaneously with a 0 or 1 planet transit model to compute the True and False Inclusion Probabilities (TIP and FIP; \citealt{Hara2022}) for the presence of transit in the detrended data. We constructed the planet transit model using priors on transit center time and orbital periods from our RV analysis, and a transit depth estimated from our $M_{p}\sin{i}$ and known mass-radius relations \citep{Otegi2020} whilst accounting for the dilution due to the host star companion and leaving the impact parameter unconstrained. 

TIP/FIP values are computed via a Bayes Evidence weighted comparison of the posterior distributions of the 0 and 1 planet fits. Out of the 95 potential transits that would have occurred during the TESS observations, we found TIP values of 0.7-0.8 for three epochs (BJD$\sim$2458720.4, 2459797.1, and 2459853.1), with the remaining 92 epochs having TIPs of $\sim$0. The three noted potential transits have depths $\sim$780\,ppm, and non-grazing impact parameters. However, it should be noted that the detrended TESS data has a white noise scatter of $\sim$290\,ppm. Therefore, it is believed that the relatively high TIPs for these transit candidate epochs are due to remaining systematics within the data and that Gl 725A\,b is likely not transiting, in agreement with the transit injection-recovery test.

\section{Discussion}\label{sec:discussion}

\subsection{A S-type planet in an M dwarf system}

 Within the Solar neighbourhood (d<10 pc) only 13 multiple stellar systems are known to host exoplanets, of which 10 contain at least one M dwarf \footnote{source: the Exoplanet Archive \url{exoplanetarchive.ipac.caltech.edu}}. All of the exoplanets detected in those systems are S-type planets. The planet Gl\,725A b joins the group of close S-type low-mass planets ($M_{p}\sin{i}<5M_{\oplus}$) in short orbital periods ($P<15$ d) whose host star is an M dwarf: Proxima Centauri b and d \citep{Anglada2016,SuarezMascareno2020,Faria2022}, GJ\,15A b \citep{Howard2014,Pinamonti2018}, and LTT 1445A b and c \citep{Winters2019,Winters2022}. Of this group, only LTT\,1445A b and c have a true mass measurement thanks to their transiting nature. 

 The formation mechanism of S-type planets is affected by the stellar multiplicity. The critical semi-major axis $a_{c}$ is a first approximation to discard planetary orbits that are unstable due to the gravitational influence of the stellar companion. We derived $a_{c}$ using the binary system orbital parameters listed in Table~\ref{table:binary_results} and including the dynamical mass ratio of Gl\,725A and B, of $\mu=0.434\pm0.003$ \citep{GaiaDR3}. Following \citealt{Holman1999}, we derived a value of \hbox{$a_{c} = 12\pm1$\,AU}. If we now consider the effect of the planet orbital inclination as in \citealt{Quarles2020}, we found that for orbital inclinations of $i=0^{\circ},30^{\circ},45^{\circ}$, the critical semi-major axis is $a_{c}=12.9\pm0.4,12.4\pm0.4,8.0\pm0.5$\,AU, respectively. From these results, the planet Gl\,725A b is too close to the host star to be affected by the stellar companion as its semi-major axis is $0.068\pm0.001$\,AU. Moreover, we can discard very long-period planets in this system. Planets with periods longer than $\sim29\,500$\,days (i.e., $a_{c} = 12.9$\,AU) are predicted to be in unstable orbits. This critical period is reduced to $\sim14\,500$ (i.e., $a_{c} = 8.0$\,AU) days for orbital inclinations of $45^{\circ}$.

The adopted planet model for Gl\,725A b assumes a circular orbit. Given the large separation between the stellar components, we can discard gravitational interaction between the stellar companion and the planet. The lack of these interactions could help to maintain the circular orbit of Gl\,725A b.

\subsection{Limits on planetary mass and radius}

Due to its likely non-transiting nature (see Sect. \ref{sec:tess}), we can only derive a minimum mass for Gl\,725A b of $M_{p}\sin{i}=2.78\pm0.35\, M_{\oplus}$. 

\begin{figure}
  \centering
  \includegraphics[width=\linewidth]{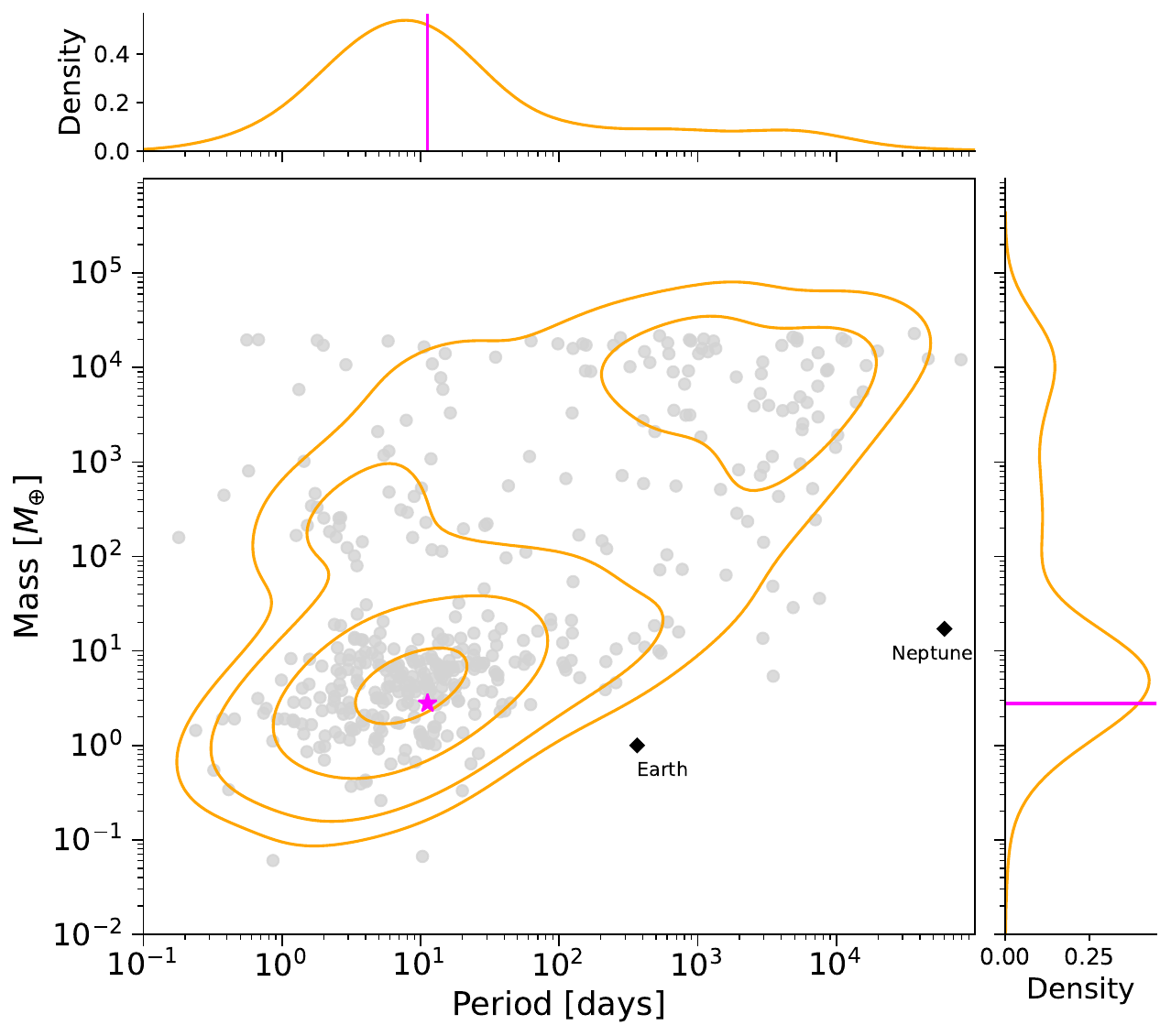}
  \caption{Orbital period versus mass diagram of planets around M dwarfs. For simplicity, planets with orbits larger than $10^5$ days are not included. The orange contours indicate 10, 30, 60, and 90$\%$ of density levels of the observed probability density of the joint mass and period distribution. The magenta star and lines indicate the position of Gl\,725A b, with a minimum mass of $M{_p}\sin{i}=2.78\pm0.35\,M_{\oplus}$, and an orbital period of 11.2\,days. The top and right histograms correspond to the probability density distribution of the orbital periods and planetary masses, respectively. For reference, the black diamonds represent the Earth and Neptune.}\label{fig:Mdwarf_planets}
   \end{figure}

The mass versus orbital period diagram for exoplanets around M dwarfs is shown in Fig. \ref{fig:Mdwarf_planets}, on which only planets with true mass measurements are included. We estimated the observed probability density of the joint mass and period distribution by means of kernel smoothing with a Gaussian kernel. The orange contours in Fig. \ref{fig:Mdwarf_planets} indicate iso-proportions of the density at the levels of 10, 30, 60, and 90$\%$. In this diagram, the estimated minimum mass for this planet falls within the innermost contour, close to the peak of both distributions. The peak of the orbital period distribution is at 7.9 days, while that of the mass distribution is at 4.9$M_{\oplus}$. In the parameter space of period versus mass, Gl\,725A b is placed in the typical type of planets discovered around M dwarfs. However, since the sample of exoplanets around M dwarfs with true mass measurements is still small (less than $10\%$ of all discoveries), this analysis can be affected by observational biases.   


Earths and super-Earths discovered orbiting M dwarfs\footnote{Confirmed exoplanets around M dwarfs, source: The Exoplanet Archive \url{exoplanetarchive.ipac.caltech.edu}} are three times more common in short-period orbits ($P<15$ d) than larger planets ($R_{\rm p}>5\,R_{\oplus}$). Moreover, in the case of M0-M3 dwarfs as Gl\,725A, \citet{Pinamonti2022} obtained a high occurrence rate of $0.85^{+0.46}_{-0.21}$ for low-mass planets ($1<M_{p}\sin{i}/M_{\oplus}<10$) in orbits between 10 and 100 days. This suggests that Gl\,725A b is likely to be a low-mass planet with a radius smaller than 5\,$\,R_{\oplus}$, as predicted using Mass-Radius relationships. A transiting planet with a radius between 1.2 and 5\,$\,R_{\oplus}$ is detectable in our current TESS data, which supports the non-transiting nature of the planet.

\subsection{Stellar activity in the optical and NIR}

Multi-wavelength observations, including optical and near-infrared data, have proven valuable in understanding stellar activity signals in low-mass stars \citep[e.g.,][]{CortesZuleta2023}. In this study, the spectropolarimetric information provided by SPIRou NIR data allowed us to accurately determine the stellar rotation period of Gl\,725A, and to confirm the planet signal independently. 

The long stellar rotation period of Gl\,725A measured thanks to the SPIRou spectropolarimetry is in agreement with the behaviour of a star with low activity levels. Moreover, from the temporal variability analysis of the optical and NIR activity indicators, we found that the rotational modulation is not significantly detected using Lomb-Scargle periodograms. Given the slow rotation rate of Gl\,725A, the distortion of the spectral lines due to activity features may be negligible, and consequently, activity proxies based on this phenomena (i.e., CCF FWHM, CCF contrast, and BIS) may not show periodicities at the stellar rotation period. The lack of detection of the rotation period in the activity indicators time series can be also explained due to the activity amplitude being smaller than the measurement precision \citep{Lafarga2021}.

The residuals of the adopted model (GP$_{\rm SOPHIE}$ + GP$_{\rm SPIRou}$ + Planet$_{b,\rm{SOPHIE+SPIRou}}$) is 2.0 m/s in the SOPHIE data and 3.2 m/s in the SPIRou data. This residual scatter level for SOPHIE is comparable with the RV uncertainties. On the other hand, it is unclear whether the higher scatter observed in the NIR is of stellar activity nature, instrumental systematics or residuals from the \texttt{wapiti} correction. 

The amplitude of the GP model applied to the stellar activity signal (see Sect. \ref{sec:keplerian}) is observed to be slightly higher for SOPHIE with a value of $2.52^{+0.67}_{-0.47}$\,m/s, in comparison with $1.43\pm0.40$\,m/s of SPIRou, but still roughly consistent within $1\sigma$. 

The RV scatter due to activity was thus found to be similar in both wavelength domains.
While M dwarfs later than M4.5 often exhibit a chromatic dependence in the amplitude of the activity-driven RV signal, \citet{Reiners2010} found that the chromaticity of the activity signal generally decreases in  earlier-type stars. This is consistent with the lack of chromaticity found in the stellar activity component of the data presented here for Gl\,725A.

\section{Summary and Conclusions}\label{sec:conclusions}

We report the discovery of the low-mass planet candidate Gl\,725A b, which orbits the brightest component of a close M dwarf binary system at 3.5 pc. Gl\,725A b has an orbit consistent with circular with a period of $11.2201\pm0.0051$ d. The RV semi-amplitude is of $1.67\pm0.20$ m/s which yields to a minimum mass of $M_{p}\sin{i}=2.78\pm0.35\,M_{\oplus}$. The discovery has been carried out with the high-precision spectrographs SOPHIE and SPIRou, in the optical and NIR wavelengths, respectively, as part of a long-term monitoring program spanning almost five years.

The measured semi-amplitude is one of the lowest detected by both instruments. In particular, for the NIR side, the signal is successfully detected thanks to the RV \texttt{wapiti} correction. The planetary feature in the RVs is consistently detected in both domains, optical and nIR, which is an important indication to confirm the planetary nature of the signal since is it not affected by chromaticity.

Using long-term astrometric data of the binary system, we updated and improved their orbital parameters of Gl\,725. We analysed TESS photometric data in search of transit events, however, the light curves are affected by blending from the stellar companion. No transit events were detected through a BLS search nor the computation of False Inclusion Probabilities. Through a injection-recovery test of transit events in the TESS light curve we conclude that Gl\,725A b is likely a non-transiting planet. Using Mass-Radius relationships, we predict a planet radius of the order of $1.4\,R_{\oplus}$, which places Gl\,725A b within the super-Earth regime.

The host star Gl\,725A is a low-activity mid-M dwarf with a long rotation period of $103.1\pm6.1$\,days. The analysis of the temporal evolution of optical and NIR activity indicators is in agreement with a less active M dwarf since rotational modulation is not detected.

This new exoplanet, likely a super-Earth, is one of the closest planets to Earth. Gl\,725A b joins the population of low-mass planets with short orbital periods (P<15 d) around close M dwarfs, which is helping us to understand better the Solar neighbourhood. This discovery also contributes to the growing body of evidence that low-mass planets are common around M dwarfs.



\begin{acknowledgements}
We thank the anonymous referee for the constructive comments and suggestions which improved the quality of this manuscript.
This research made use of Lightkurve, a Python package for Kepler and TESS data analysis \citep{lightkurve}.
This research has made use of the Washington Double Star Catalog maintained at the U.S. Naval Observatory.
This work is based on observations collected with the SOPHIE spectrograph on the 1.93\,m telescope at the Observatoire de Haute-Provence (CNRS), France. We thank the staff of the Observatoire de Haute-Provence for their support at the 1.93 m telescope and on SOPHIE.
This work is  based on observations obtained at the Canada-France-Hawaii Telescope (CFHT) which is operated from the summit of Maunakea by the National Research Council of Canada, the Institut National des Sciences de l'Univers of the Centre National de la Recherche Scientifique of France, and the University of Hawaii. The observations at the Canada-France-Hawaii Telescope were performed with care and respect from the summit of Maunakea which is a significant cultural and historic site. The authors wish to recognize and acknowledge the very significant cultural role and reverence that the summit of MaunaKea has always had within the indigenous Hawaiian community. We are most fortunate to have the opportunity to conduct observations from this mountain.
Based on observations obtained with SPIRou, an international project led by Institut de Recherche en Astrophysique et Plan\'etologie, Toulouse, France.
This paper includes data collected by the TESS mission that are publicly available from the Mikulski Archive for Space Telescopes (MAST). 
We acknowledge funding from the European Research Council (ERC) under the H2020 research \& innovation program (grant agreement $\#$740651 NewWorlds). 
We acknowledge funding from the French National Research Agency (ANR) under contract number ANR-18-CE31-0019 (SPlaSH). 
PCZ and ACC acknowledge support from STFC consolidated grant number ST/V000861/1, and UKSA grant number ST/X002217/1. 
TWi acknowledges support from the UKSA and the University of Warwick. 
This work is supported by the French National Research Agency in the framework of the Investissements d'Avenir program (ANR-15-IDEX-02), through the funding of the ``Origin of Life" project of the Grenoble-Alpes University. SS acknowledges support from the “Programme National de Physique Stellaire“ (PNPS) and “Programme National de Planétologie“ (PNP) of CNRS/INSU co-funded by CEA and CNES. 
ÉA, CC \& NJC acknowledge the financial support of the FRQ-NT through the Centre de recherche en astrophysique du Québec as well as the support from the Trottier Family Foundation and the Trottier Institute for Research on Exoplanets.
N.A-D. acknowledges the support of FONDECYT project 1240916. This project is funded/Co-funded by the European Union (ERC, ESCAPE, project No 101044152). Views and opinions expressed are however those of the author(s) only and do not necessarily reflect those of the European Union or the European Research Council Executive Agency. Neither the European Union nor the granting authority can be held responsible for them. 
J.H.C.M. is supported by FCT - Fundação para a Ciência e a Tecnologia through national funds by these grants: UIDB/04434/2020, UIDP/04434/2020, PTDC/FIS-AST/4862/2020. J.H.C.M. is also supported by the European Union (ERC, FIERCE, 101052347). Views and opinions expressed are however those of the author(s) only and do not necessarily reflect those of the European Union or the European Research Council. Neither the European Union nor the granting authority can be held responsible for them. 
E.M. acknowledges funding from FAPEMIG under project number APQ-02493-22 and a research productivity grant number 309829/2022-4 awarded by the CNPq, Brazil.
\end{acknowledgements}

\bibliographystyle{aa}
\bibliography{references}

\begin{appendix}
\onecolumn
\FloatBarrier

\section{Log of observations}\label{App:RVs}
The log of the SOPHIE and SPIRou observations with their corresponding activity indicators are listed in Tables \ref{table:SOPHIEdata} and \ref{table:SPIRoudata}, respectively.\\

\FloatBarrier
     
\begin{longtable}{ccccccccccc} 
\caption{SOPHIE RVs and activity indicators: BIS, H$\alpha$, S index, CCF FWHM, and CCF contrast. The reported RVs are corrected by the BERV, secular acceleration, and long-term drift due to the stellar companion. Values of 999.990 were not considered in the analysis and correspond to outliers.}\\        
\label{table:SOPHIEdata} \\
\hline
BJD-2400000 & $\Delta$RV & $\sigma_{\rm RV}$&  BIS &  $\sigma_{\rm BIS}$ &   H$\alpha$ & $\sigma_{\rm H\alpha}$ &  S index & $\sigma_{\rm S}$ &  CCF FWHM & CCF contrast \\ 
d & m\,s$^{-1}$ &  m\,s$^{-1}$ & m\,s$^{-1}$ & m\,s$^{-1}$ & & & & & km\,s$^{-1}$ & $\%$ CCF \\
\hline
\endfirsthead
\caption{SOPHIE RVs and activity indicators continued.}\\
\hline
BJD-2400000 & $\Delta$RV & $\sigma_{\rm RV}$&  BIS &  $\sigma_{\rm BIS}$ &   H$\alpha$ & $\sigma_{\rm H\alpha}$ &  S index & $\sigma_{\rm S}$ &  CCF FWHM & CCF contrast \\ 
d & m\,s$^{-1}$ &  m\,s$^{-1}$ & m\,s$^{-1}$ & m\,s$^{-1}$ & & & & & km\,s$^{-1}$ & $\%$ CCF \\
\hline \hline
\endhead
\hline
\endfoot
59303.6284 &   6.8 &    1.4 &       6.1 &        2.0 &   0.352 &   0.001 &   0.640 &  0.008 & 4.508 &      28.3 \\ 59304.6404 &   2.0 &    1.4 &       8.6 &        2.0 &   0.356 &   0.001 &   0.633 &  0.007 & 4.524 &      28.3 \\ 59305.6368 &   2.0 &    1.4 &       8.4 &        2.1 &   0.346 &   0.001 &   0.586 &  0.007 & 4.523 &      28.4 \\ 59306.6492 &   0.3 &    1.5 &       9.2 &        2.2 &   0.351 &   0.001 &   0.635 &  0.008 & 4.523 &      28.1 \\ 59328.6184 &   0.9 &    1.5 &       6.2 &        2.2 &   0.360 &   0.001 &   0.614 &  0.008 & 4.516 &      28.0 \\ 59342.5254 &   2.5 &    1.7 &       3.8 &        2.4 &   0.358 &   0.001 &   0.617 &  0.010 & 4.522 &      27.7 \\ 59343.4874 &  -0.3 &    1.9 &       8.2 &        2.8 &   0.361 &   0.001 &   0.749 &  0.014 & 4.506 &      26.6 \\ 59359.4885 &   6.3 &    1.7 &       7.9 &        2.4 &   0.354 &   0.001 &   0.585 &  0.009 & 4.516 &      27.6 \\ 59362.6085 &   0.1 &    1.3 &       7.5 &        1.8 &   0.356 &   0.001 &   0.574 &  0.006 & 4.525 &      28.7 \\ 59368.5240 &   2.7 &    1.4 &       6.8 &        2.0 &   0.360 &   0.001 &   0.619 &  0.007 & 4.522 &      28.6 \\ 59374.5448 &  -3.7 &    1.7 &       2.8 &        2.4 &   0.358 &   0.001 &   0.612 &  0.009 & 4.519 &      27.9 \\ 59376.5103 &   0.8 &    1.6 &       6.2 &        2.2 &   0.354 &   0.001 &   0.624 &  0.009 & 4.509 &      28.1 \\ 59379.4990 &   0.9 &    1.5 &       6.6 &        2.2 &   0.354 &   0.001 &   0.630 &  0.009 & 4.514 &      28.2 \\ 59387.5798 &   4.0 &    2.3 &       0.3 &        3.3 &   0.352 &   0.002 &   0.835 &  0.016 & 4.504 &      25.7 \\ 59392.5379 &   2.6 &    1.8 &       7.3 &        2.5 &   0.350 &   0.001 &   0.661 &  0.010 & 4.513 &      27.3 \\ 59398.4552 &  -1.5 &    1.9 &       6.3 &        2.7 &   0.367 &   0.001 &   0.605 &  0.012 & 4.515 &      27.1 \\ 59405.5215 &   1.4 &    1.4 &       8.3 &        2.0 &   0.364 &   0.001 &   0.596 &  0.007 & 4.518 &      28.6 \\ 59407.4875 &  -5.9 &    2.9 &       0.7 &        4.2 &   0.360 &   0.002 &   0.934 &  0.022 & 4.511 &      23.3 \\ 59411.5010 &   3.4 &    1.6 &       6.7 &        2.3 &   0.364 &   0.001 &   0.631 &  0.008 & 4.519 &      27.6 \\ 59413.4793 &   2.7 &    1.4 &       9.0 &        1.9 &   0.366 &   0.001 &   0.611 &  0.006 & 4.514 &      28.7 \\ 59418.4138 &  -2.0 &    1.6 &      10.3 &        2.4 &   0.375 &   0.001 &   0.653 &  0.010 & 4.498 &      27.6 \\ 59421.4742 &  -4.1 &    1.7 &       4.9 &        2.4 &   0.378 &   0.001 &   0.648 &  0.010 & 4.499 &      27.6 \\ 59425.5281 &  -2.8 &    1.7 &       7.6 &        2.4 &   0.380 &   0.001 &   0.664 &  0.010 & 4.492 &      27.5 \\ 59432.4387 &   0.7 &    2.1 &      -3.0 &        3.0 &   0.364 &   0.001 &   0.724 &  0.013 & 4.512 &      26.4 \\ 59435.4794 &  -0.9 &    1.6 &       4.5 &        2.4 &   0.367 &   0.001 &   0.679 &  0.009 & 4.516 &      27.7 \\ 59443.4960 &  -0.8 &    3.1 &       8.3 &        4.4 &   0.375 &   0.002 &   0.914 &  0.026 & 4.525 &      22.6 \\ 59449.4346 &  -0.7 &    1.8 &      -0.7 &        2.6 &   0.361 &   0.001 &   0.688 &  0.012 & 4.506 &      27.1 \\ 59451.4991 &  -4.8 &    1.9 &      -0.0 &        2.7 &   0.362 &   0.001 &   0.777 &  0.015 & 4.482 &      26.8 \\ 59454.4516 &  -1.3 &    1.9 &      -5.2 &        2.7 &   0.355 &   0.001 &   0.712 &  0.012 & 4.511 &      26.9 \\ 59457.4454 &   4.6 &    1.5 &       7.8 &        2.2 &   0.348 &   0.001 &   0.696 &  0.008 & 4.516 &      28.0 \\ 59463.3335 &   0.1 &    1.6 &       3.0 &        2.3 &   0.345 &   0.001 &   0.681 &  0.009 & 4.513 &      27.9 \\ 59468.3440 &   6.0 &    1.3 &       7.3 &        1.8 &   0.348 &   0.001 &   0.658 &  0.006 & 4.518 &      28.7 \\ 59470.4531 &   1.3 &    1.6 &       2.2 &        2.3 &   0.349 &   0.001 &   0.811 &  0.010 & 4.506 &      27.5 \\ 59477.3879 &  -9.1 &    2.8 &      -4.3 &        4.0 &   0.349 &   0.002 &   0.830 &  0.022 & 4.519 &      23.9 \\ 59485.3871 &  -3.4 &    1.8 &       6.4 &        2.6 &   0.336 &   0.001 &   0.762 &  0.011 & 4.518 &      27.3 \\ 59488.3988 &   2.1 &    1.3 &       6.2 &        1.9 &   0.332 &   0.001 &   0.705 &  0.007 & 4.508 &      28.7 \\ 59501.2978 &   0.3 &    1.7 &       6.5 &        2.4 &   0.356 &   0.001 &   0.689 &  0.010 & 4.513 &      27.6 \\ 59503.3657 &   1.3 &    1.5 &       5.5 &        2.2 &   0.360 &   0.001 &   0.736 &  0.010 & 4.498 &      27.9 \\ 59505.3043 &  -3.5 &    1.5 &       5.5 &        2.2 &   0.358 &   0.001 &   0.655 &  0.008 & 4.513 &      28.2 \\ 59507.3035 &  -0.9 &    1.6 &       4.8 &        2.3 &   0.364 &   0.001 &   0.719 &  0.009 & 4.507 &      27.9 \\ 59512.2596 &   2.4 &    1.4 &       4.0 &        2.0 &   0.362 &   0.001 &   0.636 &  0.007 & 4.511 &      28.6 \\ 59514.2844 &   4.9 &    1.4 &       3.2 &        2.0 &   0.363 &   0.001 &   0.610 &  0.007 & 4.514 &      28.7 \\ 59520.2527 &  -0.9 &    1.8 &      -1.7 &        2.6 & 999.990 & 999.990 & 999.990 &  999.990 & 4.503 &      27.4 \\ 59524.2421 &   4.1 &    1.4 &      -1.1 &        2.0 &   0.367 &   0.001 &   0.630 &  0.007 & 4.509 &      28.4 \\ 59527.3459 &   3.7 &    1.7 &       1.7 &        2.5 &   0.377 &   0.001 &   1.047 &  0.017 & 4.440 &      26.7 \\ 59532.2510 &   4.4 &    1.7 &       1.9 &        2.4 &   0.382 &   0.001 &   0.729 &  0.011 & 4.495 &      27.5 \\ 59712.5126 &  -5.1 &    2.3 &       2.5 &        3.3 &   0.354 &   0.001 &   0.676 &  0.015 & 4.510 &      28.2 \\ 59722.5668 &  -0.1 &    1.9 &      10.1 &        2.7 &   0.344 &   0.001 &   0.637 &  0.011 & 4.505 &      28.7 \\ 59728.5983 &  -0.6 &    2.2 &       8.1 &        3.1 &   0.343 &   0.001 &   0.797 &  0.014 & 4.506 &      28.3 \\ 59732.5703 &  -1.0 &    1.9 &       4.8 &        2.7 &   0.347 &   0.001 &   0.643 &  0.011 & 4.515 &      28.9 \\ 59736.5039 &   3.8 &    1.8 &       7.1 &        2.5 &   0.342 &   0.001 &   0.663 &  0.010 & 4.513 &      29.0 \\ 59738.5623 &   0.6 &    2.6 &       0.2 &        3.7 & 999.990 & 999.990 & 999.990 &  999.990 & 4.516 &      27.8 \\ 59744.5103 &  -1.2 &    2.6 &       5.9 &        3.6 &   0.346 &   0.002 &   0.794 &  0.018 & 4.528 &      27.7 \\ 59747.5840 &  -0.4 &    2.3 &      10.8 &        3.2 &   0.431 &   0.002 &   1.341 &  0.015 & 4.533 &      28.2 \\ 59750.4467 &   1.7 &    2.1 &       5.1 &        3.0 &   0.356 &   0.001 &   0.766 &  0.014 & 4.519 &      28.5 \\ 59774.5409 &  -3.2 &    1.7 &       7.0 &        2.4 &   0.347 &   0.001 &   0.684 &  0.010 & 4.509 &      28.9 \\ 59777.4971 &  -4.9 &    2.0 &       7.4 &        2.9 &   0.347 &   0.001 &   0.760 &  0.012 & 4.518 &      28.5 \\ 59779.4406 &  -3.7 &    1.8 &       4.4 &        2.6 &   0.346 &   0.001 &   0.728 &  0.012 & 4.492 &      28.6 \\ 59784.3909 &  -0.2 &    2.0 &      11.2 &        2.8 &   0.362 &   0.001 &   0.703 &  0.013 & 4.498 &      28.6 \\ 59799.4436 &  -3.5 &    1.9 &       6.0 &        2.8 &   0.368 &   0.001 &   0.645 &  0.012 & 4.501 &      28.5 \\ 59801.4705 &  -2.8 &    1.9 &       8.5 &        2.7 &   0.358 &   0.001 &   0.705 &  0.012 & 4.493 &      28.5 \\ 59805.5566 &  -7.1 &    2.5 &       0.7 &        3.6 &   0.372 &   0.002 &   0.821 &  0.022 & 4.476 &      27.3 \\ 59808.3948 &  -2.5 &    1.7 &       5.2 &        2.4 &   0.365 &   0.001 &   0.624 &  0.009 & 4.497 &      29.0 \\ 59811.4441 &  -5.6 &    2.3 &       4.2 &        3.3 &   0.372 &   0.001 &   0.629 &  0.016 & 4.507 &      28.1 \\ 59813.3479 &  -4.3 &    2.1 &      -0.6 &        2.9 &   0.374 &   0.001 &   0.643 &  0.013 & 4.505 &      28.5 \\ 59816.3732 &  -1.9 &    1.7 &       7.1 &        2.4 &   0.370 &   0.001 &   0.611 &  0.009 & 4.502 &      29.0 \\ 59824.4539 &  -0.7 &    1.6 &       6.1 &        2.3 &   0.355 &   0.001 &   0.561 &  0.010 & 4.492 &      29.0 \\ 59827.3669 &   0.7 &    1.7 &       9.3 &        2.4 &   0.376 &   0.001 &   0.588 &  0.010 & 4.503 &      29.0 \\ 59828.3534 &   1.6 &    1.6 &       4.5 &        2.3 &   0.368 &   0.001 &   0.585 &  0.008 & 4.512 &      29.2 \\ 59833.3970 &  -0.9 &    2.0 &       8.9 &        2.9 &   0.366 &   0.001 &   0.662 &  0.013 & 4.510 &      28.5 \\ 59835.3997 &  -1.2 &    1.7 &       6.7 &        2.4 &   0.361 &   0.001 &   0.580 &  0.009 & 4.514 &      29.2 \\ 59841.3689 &  -1.2 &    2.0 &       2.3 &        2.8 &   0.362 &   0.001 &   0.865 &  0.011 & 4.509 &      28.0 \\ 59855.3499 &  -4.9 &    1.7 &       7.1 &        2.4 &   0.336 &   0.001 &   0.674 &  0.010 & 4.511 &      29.0 \\ 59866.4344 &  -5.6 &    2.4 &       2.3 &        3.3 &   0.373 &   0.001 &   1.161 &  0.041 & 4.430 &      26.8 \\ 59868.3551 &   1.0 &    1.4 &       2.4 &        2.0 &   0.356 &   0.001 &   0.672 &  0.008 & 4.510 &      29.3 \\ 59869.2934 &  -0.0 &    1.5 &      -0.7 &        2.2 &   0.351 &   0.001 &   0.714 &  0.009 & 4.513 &      29.1 \\ 59870.2656 &   1.5 &    1.5 &       0.6 &        2.2 &   0.357 &   0.001 &   0.688 &  0.008 & 4.521 &      29.2 \\ 59881.3261 &   0.7 &    2.2 &       3.4 &        3.1 &   0.353 &   0.001 &   0.740 &  0.016 & 4.513 &      28.3 \\ 59882.3223 &  -2.6 &    2.1 &       1.2 &        2.9 &   0.364 &   0.001 &   0.788 &  0.015 & 4.507 &      28.5 \\ 59885.2859 &   8.0 &    1.9 &       6.8 &        2.7 &   0.351 &   0.001 &   0.623 &  0.013 & 4.514 &      29.3 \\ 59886.3633 &   1.9 &    1.4 &       6.0 &        2.0 &   0.354 &   0.001 &   1.074 &  0.022 & 4.410 &      27.9 \\ 59890.2374 &   2.6 &    1.3 &       3.9 &        1.8 &   0.358 &   0.001 &   0.599 &  0.006 & 4.520 &      29.6 \\ 59891.2332 &   3.0 &    1.3 &       1.7 &        1.9 &   0.357 &   0.001 &   0.588 &  0.006 & 4.523 &      29.5 \\ 59988.7225 &   4.0 &    1.5 &       9.1 &        2.1 &   0.331 &   0.001 &   0.719 &  0.009 & 4.495 &      29.1 \\ 60007.6942 &   5.0 &    1.7 &      10.6 &        2.4 &   0.335 &   0.001 &   0.677 &  0.011 & 4.492 &      28.8 \\ 60015.6750 &   0.6 &    2.1 &       4.0 &        3.0 &   0.336 &   0.001 &   0.716 &  0.014 & 4.510 &      28.4 \\ 60042.6232 &   1.3 &    1.8 &      11.3 &        2.6 &   0.363 &   0.001 &   0.591 &  0.011 & 4.510 &      28.8 \\ 60045.6269 &  -2.1 &    2.1 &      15.3 &        3.0 &   0.358 &   0.001 &   0.635 &  0.013 & 4.514 &      28.4 \\ 60073.6275 &  -2.6 &    1.8 &       8.2 &        2.6 &   0.361 &   0.001 &   0.671 &  0.010 & 4.517 &      28.7 \\ 60076.5440 &  -5.3 &    2.6 &       4.2 &        3.7 &   0.366 &   0.002 &   0.698 &  0.019 & 4.508 &      27.8 \\ 60079.5266 &  -3.8 &    1.9 &       9.3 &        2.7 &   0.360 &   0.001 &   0.630 &  0.012 & 4.510 &      28.7 \\ 60086.5906 &   1.7 &    1.6 &       6.2 &        2.3 &   0.355 &   0.001 &   0.633 &  0.009 & 4.499 &      29.1 \\ 60088.5527 &  -1.1 &    1.9 &       9.3 &        2.7 &   0.366 &   0.001 &   0.640 &  0.011 & 4.506 &      28.8 \\ 60090.4779 &  -2.4 &    1.7 &       5.5 &        2.4 &   0.360 &   0.001 &   0.637 &  0.009 & 4.507 &      29.0 \\ 60091.4819 &  -3.3 &    1.6 &       7.8 &        2.3 &   0.359 &   0.001 &   0.611 &  0.009 & 4.506 &      29.1 \\ 60097.5925 &  -1.4 &    1.6 &       3.4 &        2.2 &   0.344 &   0.001 &   0.641 &  0.009 & 4.504 &      29.2 \\ 60098.5249 &  -2.4 &    1.6 &       4.4 &        2.3 &   0.346 &   0.001 &   0.603 &  0.008 & 4.507 &      29.2 \\ 60106.5114 &   3.3 &    1.5 &       7.4 &        2.1 &   0.345 &   0.001 &   0.637 &  0.007 & 4.509 &      29.5 \\ 60111.4514 &   2.6 &    1.3 &       7.8 &        1.9 &   0.344 &   0.001 &   0.636 &  0.007 & 4.485 &      29.4 \\ 60112.4853 &   1.0 &    1.2 &       5.8 &        1.7 &   0.334 &   0.001 &   0.622 &  0.005 & 4.492 &      29.5 \\ 60113.5962 &  -4.3 &    1.9 &       7.5 &        2.7 &   0.344 &   0.001 &   0.727 &  0.012 & 4.505 &      28.7 \\ 60115.5860 &   0.6 &    2.7 &       3.1 &        3.8 &   0.355 &   0.002 &   0.727 &  0.021 & 4.504 &      28.1 \\ 60116.4447 &  -1.2 &    2.4 &       7.5 &        3.4 &   0.342 &   0.001 &   0.730 &  0.019 & 4.504 &      28.4 \\ 60120.5017 &   4.2 &    1.5 &       7.4 &        2.2 &   0.346 &   0.001 &   0.619 &  0.008 & 4.506 &      29.3 \\ 60121.4637 &   1.2 &    1.8 &       6.5 &        2.6 &   0.347 &   0.001 &   0.670 &  0.011 & 4.503 &      29.0 \\ 60124.5071 &   2.1 &    1.8 &       3.8 &        2.6 &   0.356 &   0.001 &   0.680 &  0.011 & 4.497 &      28.8 \\ 60127.4295 &   4.5 &    1.6 &      10.6 &        2.2 &   0.357 &   0.001 &   0.584 &  0.007 & 4.517 &      29.5 \\ 60128.4706 &   1.5 &    1.8 &       6.4 &        2.6 &   0.365 &   0.001 &   0.623 &  0.010 & 4.520 &      29.0 \\ 60129.4707 &   4.5 &    1.8 &      10.8 &        2.5 & 999.990 & 999.990 & 999.990 &  999.990 & 4.516 &      29.1 \\ 60130.5014 &   1.9 &    1.5 &       9.6 &        2.1 &   0.361 &   0.001 &   0.574 &  0.007 & 4.519 &      29.5 \\ 60131.5073 &   3.2 &    1.6 &      10.6 &        2.4 &   0.357 &   0.001 &   0.603 &  0.008 & 4.514 &      29.4 \\ 60134.4450 &  -1.7 &    1.5 &      11.6 &        2.2 &   0.358 &   0.001 &   0.590 &  0.007 & 4.515 &      29.5 \\ 60135.4751 &  -0.4 &    1.6 &       7.2 &        2.2 &   0.355 &   0.001 &   0.609 &  0.008 & 4.504 &      29.4 \\ 60136.4151 &  -1.6 &    1.5 &       7.2 &        2.1 &   0.365 &   0.001 &   0.636 &  0.007 & 4.514 &      29.6 \\ 60137.4329 &  -5.3 &    2.5 &       1.8 &        3.5 &   0.363 &   0.002 &   0.748 &  0.017 & 4.508 &      28.5 \\ 60139.4773 &   3.4 &    1.3 &       5.8 &        1.9 &   0.354 &   0.001 &   0.622 &  0.006 & 4.518 &      29.4 \\ 60140.4869 &  -0.6 &    1.6 &      12.1 &        2.2 &   0.357 &   0.001 &   0.581 &  0.008 & 4.515 &      29.5 \\ 60142.4723 &   1.9 &    1.3 &      10.8 &        1.9 &   0.357 &   0.001 &   0.600 &  0.006 & 4.513 &      29.6 \\ 60146.4802 &  -2.3 &    1.5 &       9.4 &        2.1 &   0.350 &   0.001 &   0.636 &  0.007 & 4.515 &      29.6 \\ 60147.5926 &  -1.9 &    2.3 &       4.4 &        3.3 &   0.363 &   0.001 &   0.761 &  0.018 & 4.483 &      28.1 \\ 60148.5171 &  -0.6 &    1.8 &       8.6 &        2.5 &   0.361 &   0.001 &   0.634 &  0.010 & 4.496 &      29.0 \\ 60149.5104 &   0.7 &    1.6 &       7.3 &        2.2 &   0.357 &   0.001 &   0.577 &  0.008 & 4.517 &      29.6 \\ 60152.4360 &  -0.2 &    2.2 &       4.9 &        3.1 &   0.358 &   0.001 &   0.660 &  0.013 & 4.522 &      28.7 \\ 60156.5782 &  -2.1 &    1.9 &       7.8 &        2.7 &   0.362 &   0.001 &   0.707 &  0.012 & 4.503 &      28.8 \\ 60158.4183 &  -2.3 &    1.1 &       9.0 &        1.6 &   0.353 &   0.001 &   0.561 &  0.004 & 4.518 &      29.9 \\ 60165.4838 &   2.2 &    1.3 &       7.7 &        1.8 &   0.356 &   0.001 &   0.638 &  0.005 & 4.519 &      29.8 \\ 60166.4067 &   0.1 &    1.4 &       7.3 &        1.9 &   0.351 &   0.001 &   0.586 &  0.006 & 4.528 &      29.7 \\ 60167.4082 &  -0.5 &    1.5 &       9.9 &        2.1 & 999.990 & 999.990 & 999.990 &  999.990 & 4.519 &      29.6 \\ 60168.4022 &   1.3 &    1.4 &       8.0 &        2.0 &   0.359 &   0.001 &   0.594 &  0.007 & 4.512 &      29.6 \\ 60169.3554 &  -2.9 &    1.6 &       8.0 &        2.2 &   0.364 &   0.001 &   0.607 &  0.008 & 4.508 &      29.4 \\ 60170.5430 &  -0.7 &    1.5 &       4.2 &        2.2 &   0.370 &   0.001 &   0.643 &  0.009 & 4.499 &      29.4 \\ 60171.5006 &  -0.1 &    1.4 &       4.7 &        2.0 &   0.365 &   0.001 &   0.607 &  0.007 & 4.505 &      29.6 \\ 60172.3840 &   0.3 &    1.6 &       6.4 &        2.3 &   0.363 &   0.001 &   0.580 &  0.008 & 4.508 &      29.4 \\ 60175.3503 &   0.8 &    1.7 &       8.4 &        2.4 &   0.371 &   0.001 &   0.583 &  0.009 & 4.512 &      29.4 \\ 60176.4628 &   3.5 &    1.4 &       6.7 &        2.0 &   0.371 &   0.001 &   0.572 &  0.007 & 4.509 &      29.7 \\ 60177.3479 &   3.5 &    1.7 &       9.4 &        2.4 &   0.373 &   0.001 &   0.589 &  0.010 & 4.510 &      29.3 \\ 60178.3566 &   0.0 &    1.8 &       4.7 &        2.6 &   0.378 &   0.001 &   0.564 &  0.011 & 4.500 &      29.2 \\ 60179.4302 &  -0.1 &    1.5 &       4.3 &        2.2 &   0.371 &   0.001 &   0.545 &  0.008 & 4.512 &      29.6 \\ 60180.4559 &  -3.9 &    1.5 &       6.7 &        2.1 &   0.372 &   0.001 &   0.556 &  0.007 & 4.509 &      29.6 \\ 60181.3764 &   0.4 &    1.5 &       0.1 &        2.1 &   0.396 &   0.001 &   0.667 &  0.008 & 4.508 &      29.6 \\ 60196.5077 &  -4.0 &    1.6 &       8.4 &        2.4 &   0.383 &   0.001 &   0.661 &  0.017 & 4.454 &      28.8 \\ 60197.3656 &  -3.0 &    1.4 &       9.1 &        2.1 &   0.387 &   0.001 &   0.507 &  0.007 & 4.501 &      29.7 \\ 60201.3773 &  -5.6 &    1.6 &       3.0 &        2.2 &   0.388 &   0.001 &   0.514 &  0.008 & 4.506 &      29.6 \\ 60202.3464 &  -1.5 &    1.5 &       3.2 &        2.2 &   0.387 &   0.001 &   0.517 &  0.008 & 4.499 &      29.6 \\ 60211.2985 &  -3.6 &    2.4 &       8.2 &        3.4 &   0.376 &   0.002 &   0.602 &  0.018 & 4.509 &      28.4 \\ 60212.3418 &   0.5 &    1.4 &       6.3 &        2.0 &   0.365 &   0.001 &   0.541 &  0.006 & 4.512 &      29.8 \\ 60213.3273 &   1.5 &    1.4 &       4.9 &        1.9 &   0.400 &   0.001 &   0.737 &  0.006 & 4.517 &      29.8 \\ 60215.2948 &   0.2 &    1.4 &       4.8 &        1.9 &   0.346 &   0.001 &   0.579 &  0.006 & 4.513 &      29.8 \\ 60217.4791 &  -2.9 &    1.8 &       4.6 &        2.5 &   0.365 &   0.001 &   0.751 &  0.018 & 4.458 &      28.6 \\ 60220.2939 &   5.0 &    1.4 &       3.4 &        2.0 &   0.365 &   0.001 &   0.567 &  0.007 & 4.517 &      29.7 \\ 60221.3274 &  -0.2 &    1.7 &       4.1 &        2.4 &   0.360 &   0.001 &   0.582 &  0.009 & 4.509 &      29.3 \\ 60224.3507 &   0.4 &    1.2 &       9.5 &        1.8 &   0.387 &   0.001 &   0.694 &  0.006 & 4.504 &      29.8 \\ 60227.3147 &  -1.2 &    1.4 &       4.5 &        1.9 &   0.488 &   0.001 &   1.046 &  0.007 & 4.517 &      29.7 \\ 60230.3009 &   2.3 &    2.1 &       5.8 &        3.0 &   0.354 &   0.001 &   0.644 &  0.014 & 4.516 &      28.8 \\ 60232.3559 &   2.7 &    1.6 &       5.1 &        2.3 &   0.364 &   0.001 &   0.672 &  0.010 & 4.510 &      29.5 \\ 60246.3042 &   4.7 &    1.5 &      -3.5 &        2.2 &   0.353 &   0.001 &   0.586 &  0.008 & 4.508 &      29.5 \\ 60249.3066 &   0.0 &    1.8 &      -6.0 &        2.5 &   0.348 &   0.001 &   0.567 &  0.011 & 4.513 &      29.3 \\ 60254.3027 &   7.7 &    1.3 &       1.6 &        1.9 &   0.355 &   0.001 &   0.617 &  0.007 & 4.499 &      29.6 \\ 60255.3029 &   5.6 &    1.6 &       1.6 &        2.2 &   0.358 &   0.001 &   0.625 &  0.009 & 4.506 &      29.5 \\ 60258.2672 &   3.7 &    1.7 &       0.1 &        2.4 &   0.347 &   0.001 &   0.590 &  0.010 & 4.511 &      29.4 \\ 60260.2510 &  -1.4 &    2.4 &       4.5 &        3.4 &   0.345 &   0.001 &   0.696 &  0.019 & 4.514 &      28.6 \\ 60263.2729 &   3.0 &    2.5 &       5.9 &        3.6 &   0.359 &   0.002 &   0.651 &  0.020 & 4.522 &      28.6 \\ 60265.2260 &   3.3 &    2.1 &       3.4 &        2.9 &   0.357 &   0.001 &   0.664 &  0.014 & 4.514 &      28.9 \\ 60268.2289 &   7.3 &    1.6 &       6.4 &        2.2 &   0.365 &   0.001 &   0.621 &  0.009 & 4.515 &      29.5 \\  \hline
\end{longtable}

\begin{longtable}{ccccccccc} 
\caption{SPIRou RVs and activity indicators: FWHM, chromatic velocity slope (CVS), and longitudinal magnetic field ($B_\ell$). The reported RVs are corrected by the BERV, secular acceleration, and long-term drift due to the stellar companion.} \\        
\label{table:SPIRoudata} \\
\hline
BJD-2450000 & $\Delta$RV & $\sigma_{\rm RV}$ & FWHM & $\sigma_{\rm FWHM}$ & CVS & $\sigma_{\rm CVS}$ & $B_\ell$ & $\sigma_{B\ell}$\\
d & m\,s$^{-1}$ &  m\,s$^{-1}$ & km\,s$^{-1}$ & km\,s$^{-1}$ & m\,s$^{-1}$ & m\,s$^{-1}$ & G & G \\ 
\hline \hline
\endfirsthead
\caption{SPIRou RVs and activity indicators continued.}\\
\hline
BJD-2450000 & $\Delta$RV & $\sigma_{\rm RV}$ & FWHM & $\sigma_{\rm FWHM}$ & CVS & $\sigma_{\rm CVS}$ & $B_\ell$ & $\sigma_{B\ell}$\\
d & m\,s$^{-1}$ &  m\,s$^{-1}$ & km\,s$^{-1}$ & km\,s$^{-1}$ & m\,s$^{-1}$ & m\,s$^{-1}$ & G & G \\ 
\hline \hline
\endhead
\hline
\endfoot
58531.159413 &  -1.5 &    1.4 & 5.592 &     0.007 &                   8.3 &                    4.7 &  -3.1 & 10.5 \\ 58589.077844 &  -9.0 &    1.5 & 5.579 &     0.007 &                 -10.4 &                    4.8 & -15.0 &  9.7 \\ 58591.071140 &  -7.3 &    1.5 & 5.569 &     0.007 &                 -11.4 &                    4.9 & -18.2 & 10.3 \\ 58592.100327 &   0.9 &    1.1 & 5.572 &     0.007 &                 -10.7 &                    3.5 & -22.9 & 10.0 \\ 58592.100327 &   0.9 &    1.1 & 5.572 &     0.007 &                 -10.7 &                    3.5 & -41.8 & 11.2 \\ 58593.111044 &   1.0 &    1.1 & 5.569 &     0.007 &                  -5.9 &                    3.4 &  -8.5 & 10.6 \\ 58593.111044 &   1.0 &    1.1 & 5.569 &     0.007 &                  -5.9 &                    3.4 & -20.8 &  9.0 \\ 58594.107475 &   0.7 &    1.1 & 5.562 &     0.008 &                  -8.6 &                    3.6 &  -6.9 & 11.1 \\ 58594.107475 &   0.7 &    1.1 & 5.562 &     0.008 &                  -8.6 &                    3.6 & -13.1 & 10.2 \\ 58595.083556 &   2.1 &    1.7 & 5.540 &     0.008 &                   2.0 &                    5.7 & -17.1 & 11.8 \\ 58596.062912 &  -4.3 &    1.6 & 5.589 &     0.008 &                 -17.2 &                    5.1 & -25.2 & 11.4 \\ 58597.062171 &   3.1 &    2.1 & 5.522 &     0.010 &                 -22.7 &                    6.2 & -11.7 & 11.5 \\ 58598.050349 &   3.5 &    1.9 & 5.534 &     0.009 &                 -23.6 &                    5.9 & -34.9 &  9.7 \\ 58599.109071 &   0.4 &    1.8 & 5.554 &     0.008 &                  -1.5 &                    5.5 & -18.2 & 13.1 \\ 58600.063660 &  -8.5 &    1.6 & 5.570 &     0.008 &                  -0.2 &                    5.1 &  -9.5 &  9.9 \\ 58618.122727 &   4.0 &    1.6 & 5.557 &     0.008 &                  -0.7 &                    4.9 & -11.3 & 11.0 \\ 58619.009257 &  -6.7 &    1.8 & 5.545 &     0.009 &                 -15.1 &                    5.5 & -16.2 & 13.0 \\ 58648.118917 &   9.8 &    1.4 & 5.577 &     0.007 &                   8.2 &                    4.5 & -10.9 &  9.1 \\ 58648.896590 &   3.0 &    1.7 & 5.538 &     0.008 &                  -9.0 &                    5.1 & -14.0 & 11.4 \\ 58649.925506 &  -6.3 &    1.7 & 5.576 &     0.007 &                 -13.5 &                    4.8 & -13.3 &  9.6 \\ 58650.927408 &  -0.7 &    1.8 & 5.525 &     0.008 &                 -25.0 &                    5.3 &  -6.8 & 12.1 \\ 58652.882280 &  -1.1 &    1.4 & 5.563 &     0.007 &                  -5.1 &                    4.6 & -13.4 &  9.3 \\ 58653.927757 &  -2.3 &    1.5 & 5.542 &     0.007 &                 -14.5 &                    4.6 & -16.1 &  8.5 \\ 58654.852019 &  -1.7 &    1.7 & 5.551 &     0.007 &                  -1.0 &                    4.8 & -10.3 & 10.1 \\ 58744.835162 &   0.1 &    1.6 & 5.541 &     0.007 &                  -1.0 &                    4.7 & -13.8 &  9.6 \\ 58745.752525 &  -6.6 &    1.8 & 5.501 &     0.008 &                  -2.2 &                    5.3 & -25.2 & 10.5 \\ 58750.768931 &  -4.2 &    1.9 & 5.517 &     0.009 &                   4.0 &                    5.4 &  -3.9 & 10.8 \\ 58751.762706 &  -9.2 &    1.4 & 5.558 &     0.007 &                   4.6 &                    4.4 &  -0.7 &  9.6 \\ 58752.755690 &  11.3 &    1.7 & 5.573 &     0.007 &                  -4.2 &                    4.7 & -17.2 & 10.4 \\ 58758.744844 &   2.7 &    1.3 & 5.583 &     0.006 &                   1.0 &                    4.2 & -16.6 & 10.0 \\ 58760.743284 &   6.8 &    1.4 & 5.600 &     0.006 &                   0.1 &                    4.3 & -10.9 & 11.6 \\ 58761.746166 &  -2.1 &    1.5 & 5.587 &     0.006 &                   0.8 &                    4.4 & -11.3 &  8.8 \\ 58762.747014 &   0.8 &    1.5 & 5.563 &     0.007 &                  11.8 &                    4.7 & -16.3 & 11.9 \\ 58764.725395 &   4.0 &    1.7 & 5.528 &     0.008 &                   8.2 &                    5.3 &  -4.4 &  9.5 \\ 58768.746699 &  -4.8 &    1.1 & 5.595 &     0.007 &                   4.3 &                    3.5 & -12.0 & 10.7 \\ 58768.746699 &  -4.8 &    1.1 & 5.595 &     0.007 &                   4.3 &                    3.5 &  14.3 & 13.0 \\ 58769.723703 &  -2.8 &    1.4 & 5.574 &     0.006 &                  11.3 &                    4.3 &  15.3 &  9.4 \\ 58770.716653 &   0.0 &    1.4 & 5.576 &     0.006 &                  -0.3 &                    4.3 &   9.7 &  9.7 \\ 58771.710418 &   4.5 &    1.5 & 5.583 &     0.007 &                  10.2 &                    4.5 &  -3.8 &  9.9 \\ 58772.724972 &   5.8 &    1.5 & 5.552 &     0.007 &                  -6.9 &                    4.7 &   9.8 & 10.6 \\ 58787.737467 &  -4.8 &    1.5 & 5.573 &     0.007 &                  -2.8 &                    4.5 & -46.5 & 11.0 \\ 58788.723721 &  -0.2 &    1.6 & 5.558 &     0.007 &                   0.3 &                    4.6 & -27.2 &  9.9 \\ 58791.720683 &  -0.8 &    1.5 & 5.544 &     0.007 &                  -8.7 &                    4.4 & -18.1 &  8.1 \\ 58796.719410 &  -9.7 &    1.3 & 5.605 &     0.007 &                  -2.7 &                    4.3 & -17.8 & 13.7 \\ 58798.717181 &   0.2 &    1.5 & 5.578 &     0.007 &                 -13.4 &                    4.9 & -21.9 &  9.6 \\ 58799.716565 &  -0.5 &    1.5 & 5.541 &     0.007 &                  10.1 &                    4.7 & -33.7 &  9.6 \\ 58800.722121 &  -2.5 &    1.5 & 5.545 &     0.007 &                  12.0 &                    4.7 & -25.3 &  9.5 \\ 58801.717127 &   1.9 &    1.8 & 5.528 &     0.008 &                   5.2 &                    5.4 & -24.3 &  9.3 \\ 58889.180342 &   2.4 &    1.3 & 5.561 &     0.007 &                   2.6 &                    4.2 & -30.6 &  9.0 \\ 58897.167661 &   9.8 &    1.4 & 5.576 &     0.006 &                   6.5 &                    4.1 & -28.1 & 10.1 \\ 58898.176414 &   1.4 &    1.4 & 5.557 &     0.006 &                   8.9 &                    3.9 & -26.1 &  8.9 \\ 58899.165024 &  -1.8 &    1.6 & 5.589 &     0.006 &                   4.3 &                    4.3 & -21.5 &  7.5 \\ 58920.145350 &   0.1 &    1.3 & 5.540 &     0.007 &                  22.3 &                    4.1 & -26.8 &  7.8 \\ 58977.094795 &  -2.3 &    1.4 & 5.573 &     0.007 &                  -8.4 &                    4.1 &  -8.4 &  8.9 \\ 58978.048680 &  -2.1 &    1.5 & 5.554 &     0.007 &                 -10.7 &                    4.3 &  -7.5 &  7.9 \\ 58979.058732 &  -2.1 &    1.5 & 5.554 &     0.007 &                  -1.7 &                    4.4 &  -5.4 &  8.9 \\ 58980.058252 &   2.6 &    1.6 & 5.538 &     0.008 &                   4.0 &                    4.8 & -36.5 & 10.0 \\ 58982.062168 &  -0.4 &    1.4 & 5.563 &     0.007 &                  -8.6 &                    4.2 & -13.0 &  7.9 \\ 58983.056478 &  -3.9 &    1.4 & 5.570 &     0.007 &                  -6.6 &                    4.1 &  -7.9 &  9.8 \\ 58984.063693 &   3.2 &    1.4 & 5.562 &     0.006 &                  -9.9 &                    4.1 & -15.0 &  8.5 \\ 58985.047943 &   3.6 &    1.3 & 5.581 &     0.006 &                 -22.4 &                    4.0 &  -7.7 &  9.9 \\ 59000.045452 &   0.7 &    1.5 & 5.549 &     0.007 &                   8.1 &                    4.5 & -10.1 &  8.8 \\ 59000.949835 &   1.9 &    1.8 & 5.505 &     0.009 &                  22.9 &                    5.3 &  -2.2 &  8.5 \\ 59003.018896 &   0.7 &    1.6 & 5.535 &     0.007 &                  16.5 &                    4.7 & -15.3 & 10.6 \\ 59004.016558 &  -0.0 &    1.5 & 5.534 &     0.007 &                  12.8 &                    4.6 & -27.0 &  9.5 \\ 59004.943200 &   0.9 &    1.4 & 5.549 &     0.007 &                   0.4 &                    4.2 & -12.8 &  7.6 \\ 59005.983694 &   3.5 &    1.2 & 5.576 &     0.006 &                  -1.1 &                    3.8 &  -6.9 &  8.7 \\ 59007.027225 &   4.0 &    1.3 & 5.549 &     0.007 &                  -6.0 &                    4.1 & -22.7 &  7.3 \\ 59007.963669 &   4.2 &    1.3 & 5.549 &     0.006 &                  -7.5 &                    4.0 & -14.5 &  7.8 \\ 59008.922292 &   1.9 &    1.6 & 5.595 &     0.007 &                  -8.8 &                    4.7 & -19.5 & 10.7 \\ 59009.967284 &   0.9 &    1.5 & 5.540 &     0.007 &                  -0.0 &                    4.3 & -32.5 &  9.2 \\ 59010.940340 &   1.1 &    1.3 & 5.563 &     0.006 &                  -5.2 &                    4.0 & -41.5 &  8.4 \\ 59030.957936 &   7.7 &    1.4 & 5.535 &     0.006 &                  -1.6 &                    4.0 &   4.2 &  7.4 \\ 59031.962240 &   6.6 &    1.3 & 5.546 &     0.006 &                 -14.0 &                    3.9 &  -1.0 &  6.6 \\ 59032.929967 &   7.4 &    1.7 & 5.493 &     0.008 &                  -5.3 &                    4.8 &  -2.7 &  7.9 \\ 59034.036255 &   0.6 &    1.6 & 5.527 &     0.007 &                  -3.1 &                    4.3 &  -0.8 &  8.5 \\ 59034.911674 &   7.9 &    1.6 & 5.522 &     0.007 &                  -7.0 &                    4.6 &   3.8 & 10.0 \\ 59035.886819 &  -5.2 &    1.4 & 5.533 &     0.007 &                  10.9 &                    4.2 &  -8.1 &  8.7 \\ 59036.936727 &  -6.4 &    1.5 & 5.530 &     0.007 &                  -9.3 &                    4.2 &   0.6 &  9.8 \\ 59037.905487 &  -3.2 &    1.7 & 5.490 &     0.008 &                   2.6 &                    4.7 & -19.4 &  8.1 \\ 59038.909207 &  -2.0 &    1.4 & 5.528 &     0.007 &                  -8.5 &                    4.1 &   0.2 &  7.6 \\ 59039.966211 &   2.7 &    1.3 & 5.564 &     0.006 &                  -2.3 &                    3.7 &   3.5 &  8.1 \\ 59040.919488 &   3.2 &    1.3 & 5.566 &     0.006 &                  -3.6 &                    4.0 &  16.7 &  9.3 \\ 59057.950063 &  -0.5 &    1.4 & 5.549 &     0.007 &                  11.4 &                    4.0 &  -8.9 &  8.2 \\ 59058.933332 &  -6.4 &    1.3 & 5.559 &     0.006 &                   4.2 &                    3.7 &  -4.1 &  8.1 \\ 59059.935371 &   0.9 &    1.4 & 5.528 &     0.007 &                   1.1 &                    4.4 & -31.1 &  8.1 \\ 59060.925991 &   0.1 &    1.5 & 5.562 &     0.007 &                  -8.9 &                    4.1 & -18.9 &  9.6 \\ 59061.911968 &  -7.4 &    1.3 & 5.585 &     0.006 &                 -11.2 &                    3.7 & -12.0 &  7.9 \\ 59062.956129 &  -5.7 &    1.3 & 5.606 &     0.006 &                   4.4 &                    3.5 & -19.8 &  8.5 \\ 59063.868568 &  -3.5 &    1.5 & 5.634 &     0.006 &                   3.5 &                    3.6 & -13.6 &  8.6 \\ 59064.985362 &  -5.1 &    1.3 & 5.608 &     0.006 &                  -7.5 &                    3.5 &  -7.8 &  8.5 \\ 59065.909669 &  -4.7 &    1.4 & 5.626 &     0.006 &                   2.5 &                    3.6 & -21.4 &  7.8 \\ 59069.895660 &  -4.2 &    1.1 & 5.598 &     0.007 &                   4.2 &                    3.7 & -14.7 & 10.9 \\ 59070.904630 &  -7.0 &    1.3 & 5.575 &     0.006 &                   7.4 &                    3.8 &  -8.7 &  8.7 \\ 59072.877094 &   5.0 &    1.2 & 5.546 &     0.006 &                  11.3 &                    3.8 & -10.6 &  8.0 \\ 59087.780332 &   4.6 &    1.3 & 5.607 &     0.007 &                  -0.5 &                    4.2 & -45.3 & 12.6 \\ 59088.781493 &   2.6 &    1.3 & 5.615 &     0.006 &                   7.9 &                    3.7 & -22.0 & 10.2 \\ 59089.786686 &  -0.5 &    1.6 & 5.555 &     0.007 &                  11.7 &                    4.3 & -24.1 &  8.1 \\ 59090.740209 &  -3.0 &    1.6 & 5.557 &     0.007 &                   9.1 &                    4.2 & -18.8 &  7.6 \\ 59091.786170 &   0.6 &    1.3 & 5.569 &     0.006 &                  11.7 &                    3.7 & -22.6 &  7.8 \\ 59092.737436 &   3.1 &    1.4 & 5.557 &     0.006 &                  12.4 &                    3.9 & -29.9 &  8.5 \\ 59093.792683 &  -1.4 &    1.3 & 5.581 &     0.006 &                  12.5 &                    3.7 & -11.5 &  7.7 \\ 59094.795488 &  -1.9 &    1.5 & 5.601 &     0.006 &                  -4.2 &                    4.0 & -17.7 &  8.4 \\ 59095.728885 &   2.7 &    1.2 & 5.586 &     0.006 &                   8.5 &                    3.5 & -16.6 &  8.2 \\ 59096.840124 &   1.5 &    1.6 & 5.524 &     0.007 &                   5.2 &                    4.2 & -24.4 & 10.5 \\ 59098.782710 &  -2.8 &    1.3 & 5.595 &     0.006 &                  12.2 &                    3.7 & -20.0 &  7.0 \\ 59099.803834 &   2.7 &    1.2 & 5.594 &     0.006 &                   7.8 &                    3.5 & -25.9 &  8.7 \\ 59100.895850 &   0.7 &    1.4 & 5.538 &     0.007 &                  11.7 &                    4.2 & -17.5 &  8.4 \\ 59101.826987 &   2.1 &    1.4 & 5.540 &     0.007 &                  12.8 &                    4.0 & -21.6 &  7.8 \\ 59102.835919 &  -0.3 &    1.3 & 5.533 &     0.006 &                  11.2 &                    3.8 & -18.8 &  9.3 \\ 59110.744358 &   5.3 &    1.5 & 5.521 &     0.007 &                   3.6 &                    4.3 & -18.4 &  9.4 \\ 59111.729098 &   5.1 &    1.6 & 5.540 &     0.007 &                  -6.7 &                    4.8 & -27.6 & 10.9 \\ 59112.717967 &  -0.5 &    1.4 & 5.586 &     0.006 &                   1.7 &                    3.6 & -29.5 &  7.4 \\ 59113.726251 &   2.4 &    1.3 & 5.615 &     0.005 &                   6.4 &                    3.5 & -29.3 &  9.1 \\ 59114.776722 &  -0.4 &    1.5 & 5.614 &     0.006 &                  19.0 &                    3.6 & -26.9 &  8.2 \\ 59115.770686 &  -0.6 &    1.4 & 5.540 &     0.007 &                  14.4 &                    3.8 & -34.7 &  8.4 \\ 59117.796994 &  -3.6 &    1.3 & 5.558 &     0.006 &                  18.6 &                    3.6 & -20.1 &  8.2 \\ 59118.770768 &  -2.7 &    1.2 & 5.583 &     0.006 &                   9.5 &                    3.5 & -31.2 &  8.7 \\ 59119.761441 &  -2.2 &    1.4 & 5.619 &     0.006 &                   3.6 &                    3.6 & -16.3 &  7.9 \\ 59120.723602 &   2.7 &    1.3 & 5.600 &     0.005 &                   9.3 &                    3.4 & -18.1 &  7.5 \\ 59121.724532 &   5.1 &    1.4 & 5.582 &     0.006 &                   0.2 &                    3.7 & -31.3 &  8.9 \\ 59122.733009 &   0.4 &    1.6 & 5.496 &     0.007 &                  18.8 &                    4.3 & -20.7 &  8.2 \\ 59123.726778 &  -0.0 &    1.6 & 5.505 &     0.007 &                   9.3 &                    4.2 & -22.1 &  8.3 \\ 59125.726250 &  -4.1 &    1.4 & 5.562 &     0.006 &                  18.5 &                    3.9 & -13.1 & 10.2 \\ 59126.733114 &  -1.6 &    1.2 & 5.572 &     0.006 &                   4.8 &                    3.4 & -14.3 &  8.3 \\ 59127.723206 &   0.8 &    1.3 & 5.561 &     0.006 &                   3.3 &                    3.6 & -21.5 &  8.2 \\ 59128.709330 &   3.1 &    1.4 & 5.566 &     0.006 &                   3.4 &                    4.1 & -17.3 & 10.3 \\ 59129.713451 &  -0.3 &    1.3 & 5.530 &     0.006 &                  17.8 &                    3.7 &  -8.4 &  7.0 \\ 59130.698769 &  -1.5 &    1.4 & 5.524 &     0.006 &                  14.6 &                    3.8 & -15.2 &  8.3 \\ 59153.700153 &  -0.1 &    1.4 & 5.520 &     0.006 &                   6.9 &                    3.6 &  -5.5 &  7.5 \\ 59154.704000 &   1.3 &    1.4 & 5.508 &     0.007 &                  18.8 &                    3.8 & -13.0 &  7.0 \\ 59158.703948 &  -6.4 &    1.4 & 5.513 &     0.007 &                  11.8 &                    4.0 & -14.6 &  7.6 \\ 59159.704939 &  -3.1 &    1.6 & 5.569 &     0.007 &                   7.1 &                    5.1 & -15.3 & 12.3 \\ 59296.131273 &  -0.1 &    1.2 & 5.552 &     0.006 &                  -3.4 &                    3.5 &  -2.8 &  6.2 \\ 59300.077737 &  -0.4 &    1.3 & 5.539 &     0.006 &                   1.9 &                    3.6 & -10.8 &  7.2 \\ 59306.106551 &  -2.2 &    1.3 & 5.574 &     0.006 &                   0.6 &                    3.8 &  -2.2 &  8.1 \\ 59327.019588 &  -2.5 &    1.4 & 5.550 &     0.007 &                  -3.5 &                    4.0 & -35.3 &  7.4 \\ 59328.112982 &  -0.0 &    1.4 & 5.548 &     0.007 &                  -3.5 &                    4.0 & -28.0 &  9.0 \\ 59329.032464 &  -3.0 &    1.3 & 5.586 &     0.006 &                   0.4 &                    3.8 &  -7.3 &  9.2 \\ 59330.026324 &  -3.4 &    1.2 & 5.572 &     0.006 &                  -2.5 &                    3.4 & -29.4 &  8.2 \\ 59332.075842 &  -0.8 &    1.5 & 5.575 &     0.007 &                  -5.4 &                    4.1 & -28.2 &  9.0 \\ 59333.077572 &   3.7 &    1.2 & 5.561 &     0.006 &                  -0.2 &                    3.5 & -38.4 &  7.0 \\ 59335.016251 &   8.7 &    1.5 & 5.537 &     0.007 &                   0.5 &                    4.3 & -39.6 &  8.3 \\ 59336.019673 &   7.1 &    1.4 & 5.532 &     0.007 &                  -0.6 &                    4.1 & -14.6 &  8.9 \\ 59337.042691 &   2.8 &    1.2 & 5.582 &     0.006 &                  -6.8 &                    3.5 & -26.0 &  7.5 \\ 59338.053839 &   4.8 &    1.4 & 5.592 &     0.006 &                  -1.3 &                    3.5 & -22.8 &  6.9 \\ 59384.910601 &  -7.5 &    1.4 & 5.554 &     0.006 &                  20.7 &                    3.9 &  -9.1 &  8.2 \\ 59385.883363 &   0.6 &    1.3 & 5.545 &     0.006 &                  -0.4 &                    3.6 & -16.3 &  8.2 \\ 59386.876509 &  -8.0 &    1.3 & 5.564 &     0.005 &                   0.9 &                    3.4 & -13.4 &  7.7 \\ 59387.912487 &  -0.1 &    1.6 & 5.562 &     0.007 &                  -7.1 &                    4.6 &  12.3 & 11.9 \\ 59388.882732 &  -2.5 &    1.3 & 5.546 &     0.006 &                  -6.7 &                    3.5 &  -8.2 &  6.8 \\ 59389.905938 &   1.8 &    1.4 & 5.513 &     0.007 &                   7.4 &                    4.0 &  -2.8 &  8.6 \\ 59390.907856 &   4.7 &    1.3 & 5.533 &     0.006 &                   5.3 &                    4.0 & -10.3 &  9.7 \\ 59391.911387 &  -1.3 &    1.5 & 5.501 &     0.007 &                 -10.5 &                    4.4 &  -1.2 &  8.3 \\ 59393.926381 &  -2.5 &    1.6 & 5.482 &     0.008 &                  -3.5 &                    4.8 & -13.1 & 11.8 \\ 59394.919589 &   0.4 &    1.3 & 5.530 &     0.006 &                   1.6 &                    3.9 &   2.6 &  8.0 \\ 59396.048585 &  -4.6 &    1.3 & 5.564 &     0.006 &                  -9.2 &                    3.5 &  -9.2 &  7.7 \\ 59396.919941 &  -0.7 &    1.2 & 5.565 &     0.006 &                   6.2 &                    3.8 &  -4.3 &  9.2 \\ 59397.925534 &  -1.2 &    1.2 & 5.561 &     0.006 &                  -4.9 &                    3.7 & -18.5 &  8.0 \\ 59412.912590 &  -3.0 &    1.3 & 5.568 &     0.006 &                  -8.0 &                    3.7 & -23.1 &  8.5 \\ 59413.873154 &  -3.3 &    1.3 & 5.559 &     0.006 &                  -8.4 &                    3.5 &  -5.5 &  7.4 \\ 59414.837331 &   0.4 &    1.2 & 5.545 &     0.006 &                  14.8 &                    3.6 &  -8.3 &  7.2 \\ 59416.887505 &  -2.7 &    1.1 & 5.578 &     0.007 &                   5.0 &                    3.3 & -27.9 & 12.9 \\ 59416.887505 &  -2.7 &    1.1 & 5.578 &     0.007 &                   5.0 &                    3.3 & -12.5 & 12.1 \\ 59420.910473 &   0.4 &    1.4 & 5.584 &     0.007 &                  18.7 &                    4.2 & -22.0 & 10.7 \\ 59422.960740 &  -2.5 &    1.2 & 5.574 &     0.005 &                   3.4 &                    3.4 & -18.8 &  7.1 \\ 59439.792322 &   2.1 &    1.3 & 5.575 &     0.006 &                   1.8 &                    3.7 & -26.4 &  7.2 \\ 59440.781650 &   2.0 &    1.5 & 5.572 &     0.007 &                   4.2 &                    4.1 & -10.6 &  7.1 \\ 59441.780849 &   0.0 &    1.3 & 5.564 &     0.006 &                  -6.7 &                    3.9 & -16.1 &  9.0 \\ 59442.761709 &   6.1 &    1.5 & 5.568 &     0.007 &                   1.9 &                    4.1 & -17.8 &  7.9 \\ 59443.781628 &   1.8 &    1.3 & 5.577 &     0.006 &                   2.8 &                    3.8 & -31.8 &  8.2 \\ 59444.931188 &   6.1 &    1.6 & 5.514 &     0.008 &                  -6.0 &                    4.5 & -16.4 &  7.2 \\ 59445.783247 &   2.8 &    1.5 & 5.631 &     0.006 &                  -1.1 &                    3.7 & -22.5 &  8.8 \\ 59446.801455 &   2.9 &    1.3 & 5.652 &     0.005 &                   1.6 &                    3.5 &  -8.4 &  9.4 \\ 59447.836533 &   6.1 &    1.3 & 5.575 &     0.006 &                  -4.2 &                    3.8 & -14.1 &  9.9 \\ 59448.823860 &   5.7 &    1.1 & 5.612 &     0.006 &                   3.4 &                    3.5 &  -6.8 &  8.8 \\ 59449.811703 &   2.6 &    1.3 & 5.551 &     0.006 &                  -4.9 &                    3.8 & -12.7 &  8.9 \\ 59450.788965 &   3.1 &    1.5 & 5.538 &     0.007 &                   6.4 &                    3.9 & -26.5 &  7.6 \\ 59451.795111 &   0.1 &    1.3 & 5.554 &     0.006 &                   3.7 &                    3.7 & -22.7 &  6.6 \\ 59452.777808 &  -0.9 &    1.3 & 5.628 &     0.005 &                   4.4 &                    3.3 &   4.6 &  6.9 \\ 59472.744998 &  -0.2 &    1.2 & 5.569 &     0.006 &                  -7.7 &                    3.4 &   2.3 &  7.2 \\ 59475.761357 &   1.1 &    1.4 & 5.609 &     0.006 &                   6.4 &                    3.3 & -11.2 &  7.0 \\ 59476.737002 &  -1.9 &    1.7 & 5.609 &     0.007 &                   6.1 &                    4.0 &  -7.3 &  7.1 \\ 59477.757093 &   0.8 &    1.1 & 5.580 &     0.005 &                  -0.7 &                    3.2 & -15.1 &  6.6 \\ 59478.740542 &  -1.0 &    1.4 & 5.577 &     0.006 &                  10.6 &                    3.6 & -11.0 &  7.5 \\ 59479.732426 &   1.2 &    1.3 & 5.552 &     0.006 &                   0.5 &                    3.8 & -16.1 &  6.9 \\ 59480.738227 &  -4.9 &    1.5 & 5.538 &     0.007 &                  16.8 &                    4.0 & -10.9 &  6.9 \\ 59481.734659 &  -0.7 &    1.5 & 5.565 &     0.007 &                   7.7 &                    4.1 & -19.7 &  7.9 \\ 59502.762639 &   6.0 &    1.6 & 5.567 &     0.007 &                   7.4 &                    3.9 & -21.6 &  7.5 \\ 59504.794763 &  -0.2 &    1.2 & 5.550 &     0.006 &                  -2.5 &                    3.4 & -24.6 &  6.0 \\ 59506.725648 &  -0.2 &    1.4 & 5.524 &     0.006 &                   5.0 &                    3.9 & -22.6 &  7.8 \\ 59507.708142 &   0.9 &    1.5 & 5.514 &     0.007 &                   8.7 &                    4.2 & -30.8 &  7.9 \\ 59509.746490 &  -4.2 &    1.3 & 5.577 &     0.006 &                   7.8 &                    3.5 & -29.4 &  8.4 \\ 59512.694636 &   2.6 &    1.2 & 5.561 &     0.006 &                   6.5 &                    3.4 & -14.4 &  6.9 \\ 59514.694475 &  -0.1 &    1.4 & 5.516 &     0.006 &                   8.4 &                    3.8 & -22.2 &  7.1 \\ 59531.698385 &  -3.0 &    1.3 & 5.592 &     0.006 &                   7.4 &                    3.5 & -27.9 &  7.4 \\ 59537.699289 &  -4.3 &    1.5 & 5.523 &     0.007 &                  14.6 &                    4.4 & -23.6 &  8.4 \\ 59538.697863 &   2.5 &    1.3 & 5.532 &     0.006 &                  11.1 &                    3.5 & -21.8 &  6.4 \\ 59539.698195 &   0.3 &    1.4 & 5.517 &     0.007 &                   6.4 &                    3.9 & -24.9 &  7.8 \\ 59689.139606 &  -4.6 &    1.3 & 5.572 &     0.006 &                  -0.1 &                    3.7 &  -5.9 &  8.0 \\ 59690.146431 &  -5.5 &    1.3 & 5.578 &     0.006 &                 -11.7 &                    3.8 & -14.7 &  9.2\\
  \hline
\end{longtable}

\newpage
\section{Binary system model corner plot}\label{App:corner_binary}

      \begin{figure}[h!]
  \centering
   \includegraphics[width=\linewidth]{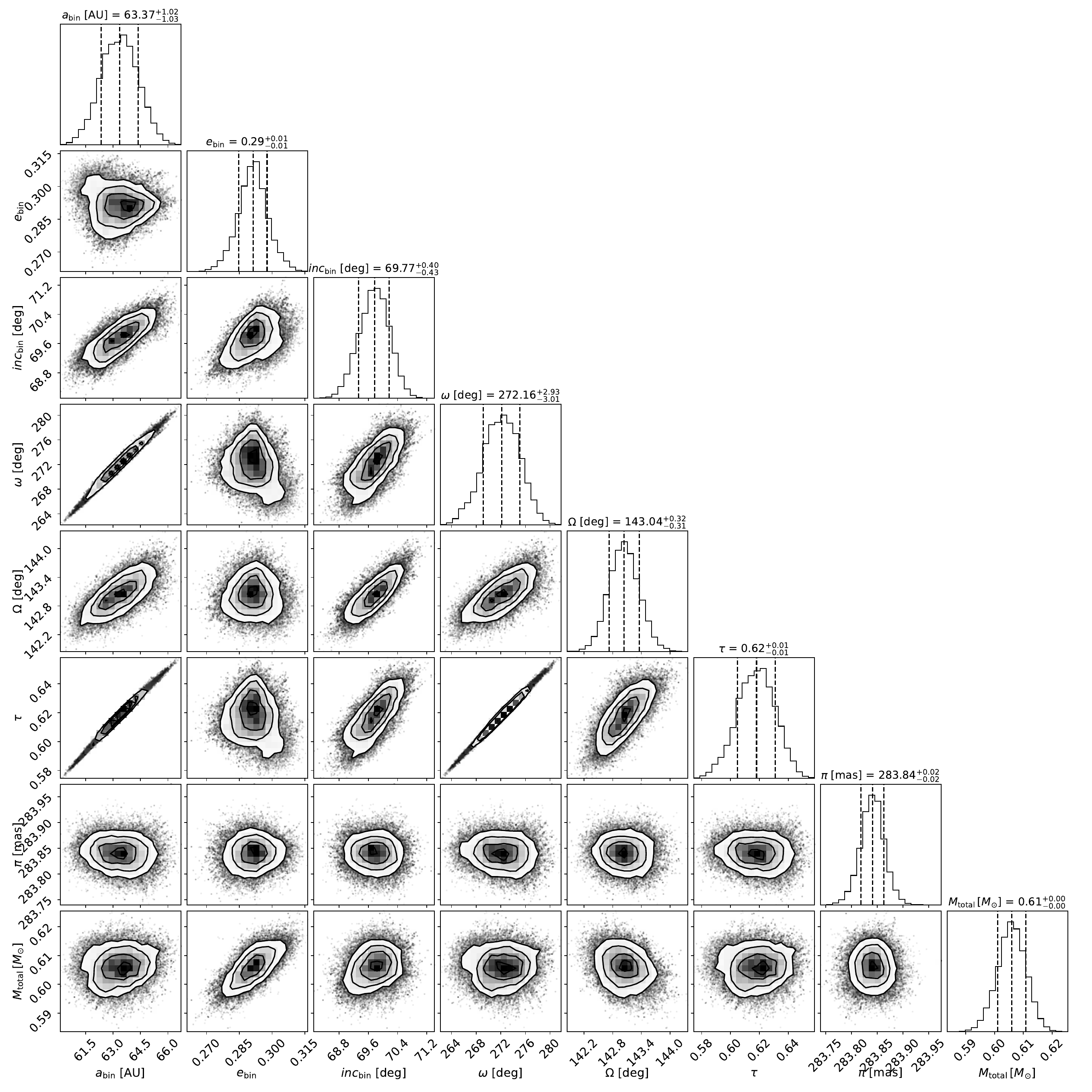}
  
  \caption{Corner plot displaying the posterior distributions of the best-fit orbital paramaters from the MCMC sampling.}\label{fig:corner_binary}
   \end{figure}
 
\newpage
\section{Planet model corner plot}\label{App:corner_planet}

  \begin{figure}[h!]
  \centering
   \includegraphics[width=\linewidth]{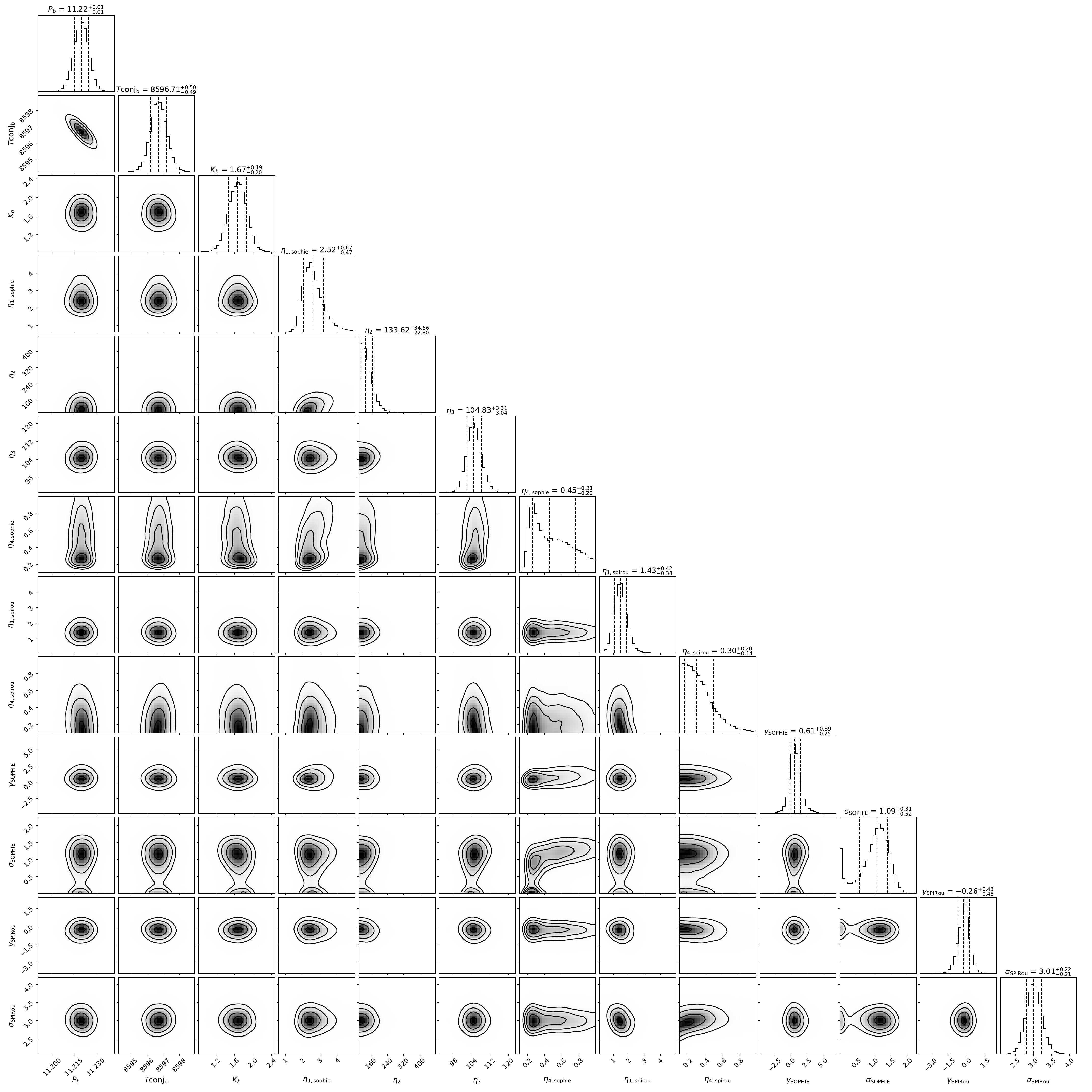}
  
  \caption{Corner plot displaying the posterior distributions of the best-fit orbital parameters and GP hyper-parameter from the MCMC sampling.}\label{fig:corner}
   \end{figure}

\end{appendix}
\end{document}